\documentclass{lmcs}
\pdfoutput=1

\usepackage{lastpage}
\lmcsdoi{22}{4}{1}
\lmcsheading{}{\pageref{LastPage}}{}{}%
{Mar.~12,~2025}{Oct.~05,~2026}{}

\pdfoutput=1


\usepackage[utf8]{inputenc}
\usepackage{sml-highlight}
\usepackage{listings}
\usepackage{xcolor}
\usepackage{graphicx}
\graphicspath{{./}}
\usepackage{proof}
\usepackage{amssymb}
\usepackage{booktabs}
\usepackage{float}
\usetikzlibrary{automata, positioning, arrows}
\tikzset{every picture/.style={line width=0.75pt}}

\newif\ifhighlightnew
\highlightnewfalse
\newcommand{\new}[1]{\ifhighlightnew\textcolor{blue}{#1}\else#1\fi}
\newcommand{\newmark}[1]{%
  \ifhighlightnew\marginpar{\scriptsize\color{blue}\textbf{[#1]}}\fi
}
\newcommand{\newid}[2]{%
  \ifhighlightnew
    \marginpar{\scriptsize\color{blue}\textbf{[#1]}}%
    \textcolor{blue}{#2}%
  \else
    #2%
  \fi
}

\newcommand{\defin}{\mathcal D}

\newcommand{\eval}[2]{#1\!\!\downarrow_{#2}}

\newcommand{\mypar}[1]{\noindent{\bf #1}}

\newcommand{\proj}{\mathsf{proj}}
\newcommand{\projD}{\proj_{\defin}}
\newcommand{\projE}[2]{#1\downarrow_{#2}}

\newcommand{\sconnected}{\mathsf{sConn}}
\newcommand{\hmodules}{\mathsf{hMods}}

\newcommand{\topannotation}[2]{\mathsf{top}_{#1}(#2)}

\definecolor{code_indent}{HTML}{CCCCCC}

\definecolor{bloodyred}{rgb}{0.6,0.1,0.1}
\newcommand{\role}[1]{\textcolor{bloodyred}{\mathtt{#1}}}
\newcommand{\nr}{\role p}
\newcommand{\nrr}{\role q}
\newcommand{\nrrr}{\role r}

\newcommand{\Var}{\mathsf{Var}}
\newcommand{\Val}{\mathsf{Val}}

\newcommand{\interactBase}[2]{{#1}\rightarrow \{{#2}\}}
\newcommand{\interactBasel}[3]{#1\rightarrow^{#2} \{#3\}}
\newcommand{\interact}[2]{\interactBase{#1}{#2}:\,\Sigma_{j\in J}\lambda_j: u_j;\ C_j}
\newcommand{\interactl}[3]{\interactBasel{#1}{#2}{#3}:\,\Sigma_{j\in J}\lambda_j: u_j;\ \anno{C}_j}

\newcommand{\allsynchName}{\textcolor{sh_keyword}{\mathtt{allsynch}}}
\newcommand{\allsynch}[4]{\allsynchName\, \{ {\role{#1}}_{#3}:#2_{#3}\}_{#3\in #4}}

\newcommand{\command}[4] {\commandBase {#1} {#2} \Sigma_{i\in I}#3_i: #4_i}
\newcommand{\commandBase}[3] {[#1]\ #2\ \rightarrow #3}
\newcommand{\chorcommandBase}[2] {#1\ \rightarrow #2}
\newcommand{\chorcommand}[3] {\chorcommandBase {#1} \Sigma_{i\in I}#2_i: #3_i}
\newcommand{\ifTE}[4]{\textcolor{sh_keyword}{\mathtt{if}}\ #1@{#2}\ \textcolor{sh_keyword}{\mathtt{then}}\ {#3} \ \textcolor{sh_keyword}{\mathtt{else}}\ {#4}}
\newcommand{\ifTEl}[5]{\textcolor{sh_keyword}{\mathtt{if}}\ #1@^{#2}{#3} \ \textcolor{sh_keyword}{\mathtt{then}}\ {#4} \ \textcolor{sh_keyword}{\mathtt{else}}\ {#5}}
\newcommand{\codeprism}[1]{\lstinline[style=prism-color,basicstyle=\small\sffamily]{#1}}
\newcommand{\codechor}[1]{\lstinline[style=chor-color,basicstyle=\small\ttfamily,mathescape=true]{#1}}

\newcommand{\CEnd}{\mathbf{0}}
\newcommand{\pp}{|\!|}
\newcommand{\ppp}[1]{|[#1]|}

\newcommand{\defrec}{\stackrel{\textcolor{sh_keyword}{\mathsf{def}}}{=}}

\newcommand{\red}[1]{\ \longrightarrow_{#1}\ } 
\newcommand{\prismred}[1]{\ \stackrel{#1}{\rightsquigarrow}\ }

\definecolor{pblue}{rgb}{0.13,0.13,1}
\definecolor{pgreen}{rgb}{0,0.5,0}
\definecolor{pred}{rgb}{0.9,0,0}
\definecolor{pgrey}{rgb}{0.46,0.45,0.48}
\definecolor{keywordColor}{rgb}{0.50,0.00,0.33}
\definecolor{stringColor}{rgb}{0.16,0.00,1.00}
\definecolor{lineNumberColor}{rgb}{0.47,0.47,0.47}

\lstdefinelanguage{Eclipse}{
language=Java,
tabsize=3,
showspaces=false,
captionpos=b,
breaklines=true,
extendedchars=true,
numbers=none,
postbreak=\raisebox{0ex}[0ex][0ex]{\ensuremath{\color{red}\hookrightarrow\space}},
  basicstyle=\ttfamily\tiny,
  emphstyle=\bfseries,
  keywordstyle=\color{keywordColor}\bfseries,
  commentstyle=\markupComments,
  stringstyle=\color{stringColor},
  numberstyle=\color{lineNumberColor}\tiny,
  morecomment=[s][\markupJavadocs]{/**}{*/}, 
  showstringspaces=false,
  morekeywords={and},
  numbers=left,
  mathescape=true
}

\begin{document}

\title[A Probabilistic Choreography Language for PRISM]
{A Probabilistic Choreography Language for PRISM}

\specialissue{Selected Papers of the 44th International Conference on Formal Techniques for Distributed Objects, Components and Systems and the 26th International Conference on Coordination Models and Languages (FORTE and COORDINATION 2024)}

\author[M.~Carbone]{Marco Carbone\lmcsorcid{0000-0001-9479-2632}}[a]
\author[A.~Veschetti]{Adele Veschetti\lmcsorcid{0000-0002-0403-1889}}[b]

\address{Computer Science Department, IT University of Copenhagen, Rued Langgaards Vej 7, 2300 Copenhagen S, Denmark}	
\email{carbonem@itu.dk, maca@itu.dk}  

\address{Department of Computer Science, TU Darmstadt, Hochschulstraße 10, 64289 Darmstadt, Germany} 
\email{adele.veschetti@tu-darmstadt.de} 

\begin{abstract}
  \noindent We present a choreographic framework for modelling and
  analysing concurrent probabilistic systems based on the PRISM
  model-checker. This is achieved through the development of a
  choreography language, which is a specification language that allows
  to describe the desired interactions within a concurrent system from
  a global viewpoint. Using choreographies provides a global view of
  system interactions, which can help in understanding process flow
  and identifying potential modelling issues.  We equip our language
  with a probabilistic semantics and then define a formal encoding
  into the PRISM language and discuss its correctness. Properties of
  programs written in our choreographic language can be model-checked
  by the PRISM model-checker via their translation into the PRISM
  language.  Finally, we implement a compiler for our language and
  demonstrate its practical applicability via examples drawn from use
  cases featured in the PRISM website.
\end{abstract}

\maketitle

\section{Introduction}\label{sec:intro}
Programming distributed systems
presents significant challenges due to their inherent complexity, and
the possibility of obscure edge cases arising from their complex
interactions.
Unlike monolithic systems, distributed programs involve multiple nodes
operating concurrently and communicating over networks, introducing a
multitude of potential failure scenarios and nondeterministic
behaviours.
One of the primary challenges in understanding distributed systems
lies in the fact that the interactions between multiple components can
diverge from the sum of their individual behaviours. This emergent
behaviour often results from subtle interactions between nodes, making
it difficult to predict and reason about a system's overall behaviour.

PRISM \cite{PRISMdoc} is a probabilistic model checker that offers a
specialised language for the specification and verification of
probabilistic concurrent systems. PRISM has been used in various
fields, including multimedia protocols \cite{multimedia}, randomised
distributed algorithms \cite{distr1,distr2}, security protocols
\cite{security1,security2}, and biological systems \cite{bio1,bio2}.
At its core, PRISM provides a declarative language with a set of
constructs for describing probabilistic behaviours and properties
within a system.
Given a distributed system, we can use PRISM to model the behaviour of
each of its nodes,
and then verify desired properties for the entire system. However,
this approach can become difficult to manage as the number of nodes
increases.

Choreographic programming~\cite{M23} is an emerging programming
paradigm for distributed systems in which programs, referred to as
choreographies, specify communication protocols from a global
perspective. Instead of programming the behaviour of each component
independently or relying on a central orchestrator, choreographic
languages describe the interactions among participants directly. A
choreography therefore abstracts away from the internal implementation
of individual components and focuses on the communication structure
that governs their coordination.
In this approach, the primary concern is not the arbitrary
interleavings of local computations but the communication structure of
the system: which participants interact, in which order, and under
which conditions. Choreographies make these interaction patterns
explicit at the level of the program.
\newid{R2.2}{Choreographic programming naturally applies to
distributed systems whose behaviour can be described through
structured interaction protocols. Systems whose dynamics rely heavily
on loosely coordinated, highly unstructured, or unpredictable
communication patterns are generally less amenable to this
approach. Within its intended scope, however, choreographic
programming has proved effective in practice, since choreographies
enable the automatic generation of decentralised implementations that
are correct by construction~\cite{CM13}.}

This paper presents a choreographic language designed for modelling
concurrent probabilistic systems and introduces a {\em projection}
function for translating protocols described in this language into
PRISM code. This choreographic approach not only simplifies the
modelling process but also ensures integration with PRISM's powerful
analysis capabilities.
In order to illustrate the idea concretely, consider the following
choreography:
\[
  \begin{array}{lll}
    C \stackrel{\mathsf{def}}{=} \interactBase{\nr}{\nrr}
    \left\{
    \begin{array}{lll}
      \lambda_1: \ (x'=1)\ \&\ (y'=2);\ C\\[1mm]
      \lambda_2: \ (x'=3)\ \&\ (y'=1);\ C
    \end{array}
    \right.
  \end{array}
\]
This specification recursively defines an interaction where process
\(\role{p}\) communicates with process \(\role{q}\), choosing the
first branch with rate $\lambda_1$ (and setting \(x\) to 1 and \(y\)
to 2) or the second branch with rate $\lambda_2$ (and setting \(x\) to
3 and \(y\) to 1). \newid{R1.4}{Here, we adopt PRISM notation where
the prime symbol, e.g., $x'$, $y'$, indicates the next-state value
of the corresponding variable, that is, the value it will take after
the transition occurs.}
The projection mechanism translates this global choreography into
local PRISM modules by annotating each branch with a unique label
(e.g., $a$ and $b$) and using a state variable (e.g., \(s_{\role p}\)
and \(s_{\role q}\)) to uniquely identify each step.
One possible representation of the behaviour described by the
choreography in PRISM is the following:
\[
\begin{array}{llll}
\role{p}: \{ & \quad \commandBase {a}{s_{\role p}=0} {\lambda_1:(x'=1)\ \&\ (s_{\role p}'=1)} \quad\qquad \commandBase {}{s_{\role p}=1} {1:(s_{\role p}'=0)}\\[2mm]
              & \quad \commandBase {b}{s_{\role p}=0} {\lambda_2:(x'=3)\ \&\ (s_{\role p}'=2)} \quad\qquad \commandBase {}{s_{\role p}=2} {1:(s_{\role p}'=0)} \quad \}\\[4mm]
\role{q}: \{ & \quad \commandBase {a}{s_{\role q}=0} {1:(y'=2)\ \&\ (s_{\role q}'=1)} \qquad\quad \commandBase {}{s_{\role q}=1} {1:(s_{\role q}'=0)}\\[2mm]
              & \quad \commandBase {b}{s_{\role q}=0} {1:(y'=1)\ \&\ (s_{\role q}'=2)} \qquad\quad \commandBase {}{s_{\role q}=2} {1:(s_{\role q}'=0)} \quad \}\\
\end{array}
\]
\newmark{R2.5}%
\new{In this translation, modules $\nr$ and $\nrr$ contain a set of
commands. For example,
$\commandBase {a}{s_{\role p}=0} {\lambda_1:(x'=1)\ \&\ (s_{\role
    p}'=1)}$ in module $\nr$ is a command where $a$ is a label over
which the module can synchronise with other modules, $s_{\role p}=0$
is a guard, $\lambda_1$ is a rate, and $(x'=1)\ \&\ (s_{\role p}'=1)$
is a list of updates for modifying the state. Here, we use labels $a$
and $b$ for simulating the two branches of the choreography.}
The command labeled \(a\) in \(\role{p}\)’s module corresponds to the
branch with rate \(\lambda_1\), and the matching command in
\(\role{q}\)’s module (with rate \(1\)) sets \(y\) to 2; similarly,
label \(b\) corresponds to the second branch with rate \(\lambda_2\)
for \(\role{p}\) and rate \(1\) for \(\role{q}\), setting \(x\) and
\(y\) appropriately.  After executing an interaction command in each
module, a subsequent reset command (with rate 1) sets the state
counters \( s_{\role p} \) and \( s_{\role q} \) back to 0 due to the
recursive call following the branching. This mechanism ensures that
these commands are only executable when the system has reached this
particular state.

Through our contributions, we aim to provide a smooth workflow for
modelling, analysing, and verifying concurrent probabilistic systems,
ultimately increasing their usability in various application domains.
In particular, by employing choreographies, we gain a clear and
comprehensive view of the interactions occurring within the system,
allowing us to discern the flow of processes and detect any potential
sources of error in the modelling phase.

\paragraph{Contributions and Overview.} We summarise our contributions as
follows:
\begin{itemize} 
\item we propose a choreographic language equipped with well-defined
  syntax and semantics, tailored specifically for describing
  concurrent systems with probabilistic behaviours
  (Section~\ref{sec:chor}); 

\item we introduce a semantics for the minimal fragment of PRISM
  needed for code generation from choreographies (Section~\ref{sec:prism}),
  which follows PRISM's original semantics~\cite{PRISMdoc};

\item we establish a rigorous definition for a translation function
  from choreographies to PRISM (Section~\ref{sec:proj}), and address its
  correctness. This translation serves as a crucial intermediary step
  in transforming models described in our choreographic language into
  PRISM-compatible representations;

\item we give an implementation~\cite{repository} that executes the
  defined translation function, enabling users to utilise PRISM's
  robust analysis features while benefiting from the expressiveness of
  choreographies;

\item we show some use cases in order to demonstrate the applicability
  of our language (Section~\ref{sec:bench}).
\end{itemize}

\paragraph{Changes with respect to the conference version.}
\newmark{R2.1}%
This article is an expanded version of our paper presented at
COORDINATION 2024~\cite{CV24}. It includes detailed definitions and a
complete proof of the main theorem \new{(Section~\ref{sec:proj})},
which were previously omitted. We also present a revised projection
function \new{(Section~\ref{sec:proj})} that provides a cleaner
formalisation and no longer relies on counters for the control state.
Furthermore, we introduce additional language constructs
\new{(Section~\ref{sub:otherconstructs})}, which are also supported by
the implementation. Additionally, we provide a more extensive discussion
of the language features supported by the implementation and expand
the related work section \new{(Section~\ref{sec:relwork})}. Finally,
we introduce new use cases analysed through choreographies
\new{(Section~\ref{sec:bench})}, including case studies on a simple
peer-to-peer protocol, synchronous leader election, and the Dining
Cryptographers protocol.

\section{A Motivating Example}\label{sec:example}
As an example, we consider a simplified version of the
\text{thinkteam} example, a three-tier data management system
\cite{DBLP:journals/entcs/BeekMLGFS05}, presented in the PRISM
documentation\footnote{\url{https://www.prismmodelchecker.org/casestudies/thinkteam.php}}.
\begin{lstlisting}[style=prism-color,% caption={A PRISM example},captionpos=b,
	frame=none,label={example1},escapechar=|]
ctmc
module User|\label{user-init}|
  User_STATE : [0..2] init 0;

  [alpha_1] (User_STATE=0) $\rightarrow$ lambda : (User_STATE'=1);|\label{first-line}|
  [alpha_2] (User_STATE=0) $\rightarrow$ lambda : (User_STATE'=2);
  [beta] (User_STATE=1) $\rightarrow$ mu : (User_STATE'=0);
  [gamma_1] (User_STATE=2) $\rightarrow$ theta : (User_STATE'=1);
  [gamma_2] (User_STATE=2) $\rightarrow$ theta : (User_STATE'=2);
endmodule|\label{user-end}|

module CheckOut|\label{check-init}|
  CheckOut_STATE : [0..1] init 0;

  [alpha_1,alpha_2] (CheckOut_STATE=0) $\rightarrow$ 1 : (CheckOut_STATE'=1);
  [beta] (CheckOut_STATE=1) $\rightarrow$ 1 : (CheckOut_STATE'=0);
  [gamma_1,gamma_2] (CheckOut_STATE=1) $\rightarrow$ 1 : (CheckOut_STATE'=1);
endmodule|\label{check-end}|
\end{lstlisting}
\noindent
The protocol aims to manage exclusive access to a single file,
  controlled by the \codeprism{CheckOut} process. Users can request
  access, which can be granted based on the file's current status. The
  goal is to ensure that only one user possesses the file at any given
  time while allowing for efficient access requests and retries in
  case of denial. Users move between different states (0, 1, or 2)
  based on the granted or denied access to the file with corresponding
  rates \new{\codeprism{lambda}, \codeprism{mu}, and
  \codeprism{theta}}\newmark{R2.6}; the \codeprism{CheckOut} process
  transitions between two states (0 or 1).

In PRISM, modules are individual processes whose behaviour is
specified by a collection of commands, in a declarative fashion.
Processes have a local state, can interact with other modules and
query each other's state. Above, the modules \codeprism{User} (lines
\ref{user-init}-\ref{user-end}) and \codeprism{CheckOut} (lines
\ref{check-init}-\ref{check-end}) can synchronise on different labels, e.g., 
\codeprism{alpha_1}. 
On line \ref{first-line}, \codeprism{(User_STATE=0)} is a condition
indicating that this transition is enabled when \codeprism{User_STATE}
has value 0. The variable \codeprism{lambda} is a rate, since the
program models a Continuous Time Markov Chain (CTMC). The command
\codeprism{(User_STATE'=1)} is an update, indicating that
\codeprism{User_STATE} changes to 1 when this transition fires.

Understanding the interactions between processes in this example may require inspecting 
several module transitions. 
When formalised using our choreographic language, 
the interaction structure is expressed as a single global description, 
which makes the communication pattern directly visible.
\begin{lstlisting}[style=chor-color,% caption={Example of Listing \ref{example1} in our choreographic language},captionpos=b,
  frame=none, label={lst:chorexample}]
C0 := CheckOut $\rightarrow$ User : (+["1*lambda"] ; C1   +["1*lambda"]  ;  C2)
C1 := CheckOut $\rightarrow$ User : (+["1*mu"] ; C0)
C2 := CheckOut $\rightarrow$ User : (+["1*theta"] ; C1   +["1*theta"] ;  C2)
\end{lstlisting}
In this model, we define three distinct choreographies, namely
\codechor{C0}, \codechor{C1}, and \codechor{C2}. These choreographies
describe the interaction patterns between the modules \codechor{CheckOut}
and \codechor{User}. The state updates resulting from these
interactions are not explicitly depicted as they are not relevant for
this particular protocol, but necessary in the PRISM implementation.
As evident from this example, the choreographic language facilitates a
straightforward understanding of the interactions between processes,
minimizing the likelihood of errors. In fact, we can think of the
choreography representation of this example as the product of the two
PRISM modules seen above.

In order to formally analyze the system, we need to project the
choreography onto individual components, resulting in a PRISM model
that captures the behaviour of each process explicitly. The projection
process ensures that each role in the choreography, such as
\codechor{User} and \codechor{CheckOut}, is assigned a state variable
that tracks its execution progress. Interactions in the choreography
correspond to PRISM synchronisation labels, ensuring consistency
across transitions. Furthermore, the rates in the choreography, such
as \new{\codeprism{lambda}, \codeprism{mu}, and \codeprism{theta}},
translate into transition rates in the PRISM model.

Each choreographic rule maps to a corresponding PRISM command. For instance, the choreographic transition 
\begin{lstlisting}[style=chor-color,frame=none,escapechar=|]
C0 := CheckOut $\rightarrow$ User : (+["1*lambda"] ; C1)
\end{lstlisting}
corresponds to the following PRISM code:
\begin{lstlisting}[style=prism-color,frame=none,escapechar=|]
$\ldots$
[alpha_1] (User_STATE=0) -> lambda : (User_STATE'=1); // User module
$\ldots$
[alpha_1] (CheckOut_STATE=0) -> 1 : (CheckOut_STATE'=1); // CheckOut module
$\ldots$
\end{lstlisting}
 This ensures both modules execute in sync under the label \codeprism{alpha_1} and only when the modules are in the correct states (\codeprism{CheckOut_STATE=0} and \codeprism{User_STATE=0}). Similar transformations apply to other interactions.

\section{Choreography Language}\label{sec:chor}
In the previous section, we illustrated our choreography language by
means of an example. We now formalise the language by presenting its
syntax and semantics. For the sake of clarity, we slightly depart from
the concrete syntax of our tool, adopting instead a more
process-algebraic format, which makes the syntax more concise and
better suited to formal definitions and proofs.

\subsection{Syntax}\label{sub:syntax}
Let $\role p$ range over a (possibly infinite) set of module names
$\mathcal R$, $x$ over a (possibly infinite) set of variables $\Var$,
and $v$ over a (possibly infinite) set of values $\Val$.
Choreographies, the key component of our language, are defined by the
following syntax:
{\small
\begin{displaymath}
  \begin{array}{llllllllllll}
    C & ::= &      & \interact{\nr}{\nr_1,\ldots,\nr_n} & \text{(interaction)}\\[1mm]
      &     & \mid & \ifTE {E}{\nr}{C_1}{C_2} & \text{(conditional)}\\[1mm]
      &     & \mid & X     & \text{(recursive call)}\\[1mm]
      &     & \mid & \CEnd & \text{(inact)}
    \\[2mm]
    \defin \ & ::= & & X\stackrel{\textcolor{sh_keyword}{\mathsf{def}}}{=} C,\ \defin 
                     \quad \mid \quad \emptyset & \text{(definitions)}
    \\[2mm]
    u     & ::=  && (x' = E)\ \&\ u \quad\mid\quad    (x' = E) & \text{(assignments)}\\[2mm]
    E, g  & ::=  && f(\tilde E)\quad\mid\quad x\quad\mid\quad v & \text{(expressions)}
  \end{array}
\end{displaymath}
}%
The syntactic category $C$ denotes choreographic programs. The
interaction term
$\interact{\nr}{\nr_1,\ldots,\nr_n}$ denotes an interaction initiated by module $\nr$ with
modules $\nr_i$'s, with $\nr$ and $\nr_i$'s all distinct. A
choreography specifies what interaction must be executed next,
shifting the focus from what can happen to what must happen. When the
interaction happens, one of the $j$ branches is selected as a
continuation. Branching is a random move: the number
$\lambda_j\in\mathbb R$ denotes either a probability or a rate. This
will depend on the language we wish to use. In the case of
probabilities, it must be the case that $0\leq\lambda_j\leq 1$ and
$\Sigma_j\lambda_j=1$. Once a branch $j$ is taken, the choreography
will execute some assignments $u_j$. A single assignment has the
syntax $(x' = E)$ meaning that the value obtained by evaluating
expression $E$ is assigned to variable $x$; assignments can be
concatenated with the operator $\&$.  Note that $x'$ is used for an
assignment to $x$: here, we follow the syntax adopted in PRISM
(see Section~\ref{sec:prism}). Expressions are obtained by applying some
unspecified functions to other expressions or, as base terms, i.e.,
variables and values (denoted by $v$).

The term $\ifTE {E}{\nr}{C_1}{C_2}$ denotes a system where module
  $\role p$ evaluates the guard $E$ (which can contain variables
  located at other modules) and then (deterministically) branches
  accordingly.  The term $X$ is a (possibly recursive) procedure call:
  in the semantics, we assume that such procedure names are defined
  separately. Moreover, we assume that such definitions are
  contractive (guarded), i.e., the body of a definition is never
  itself a procedure call: this excludes definitions such as
  $X\stackrel{\textcolor{sh_keyword}{\mathsf{def}}}{=}X$. The term
  $\CEnd$ denotes the system finishing its computation.

  \newmark{R1.11}%
  \begin{asm}\label{ass:0}
    \new{We assume that every well-formed choreography is such that, for
    every interaction $\interact{\nr}{\nr_1,\ldots,\nr_n}$, all
    updates in each $u_j$ are assignments to variables located at
    $\nr, \nr_1,\ldots,\nr_n$.}
  \end{asm}
  
\subsection{Semantics.}
The semantics of a choreography is a relation that captures how the
values assigned to variables are modified when the various modules
synchronise with each other. Similar to the operational semantics of
imperative languages, we define a state, denoted by $S$, as a mapping
from variables to values, that is, $S : \Var \rightarrow \Val$.

Given a state, substitution allows us to modify some of its values.

\newmark{R2.8}%
\begin{defi}
  \new{Given distinct variables $x_1,\ldots,x_n$ and expressions
  $E_1,\ldots,E_n$, a state update
  $S[x_1' = E_1\ \&\ \cdots \&\ x_n' = E_n]$ is defined as}
  \[
    S[x_1' = E_1\ \&\ \cdots \&\ x_n' = E_n](y) =
    \begin{cases}
      \eval{E_i}{S} & \text{if } y = x_i \text{ for some } i \in \{1, \ldots, n\},\\
      S(y)          & \text{otherwise.}
    \end{cases}
  \]
  \new{where $\eval {E_i} S$ denotes a (decidable) evaluation of the
  expression $E$ in the state $S$.}
\end{defi}

\noindent
Let $\mathcal S$ be the set of all possible states and let $\defin$ be
a set of definitions of the form
$X \stackrel{\textcolor{sh_keyword}{\mathsf{def}}}{=} C$. The
operational semantics of choreographies is defined as the smallest
relation
\[
  \red{}\!\!\!\!^{\defin} \;\subseteq\; \mathcal S \times C \times
  \mathbb R \times \mathcal S \times C
\]
satisfying the rules below (we omit $\defin$ when it is not relevant):
{\small
\begin{displaymath}
  \begin{array}{l@{\quad}llll}
    \textsf{(Interact)} &
                          (S, \interact{p}{\role p_1,\ldots,\role p_n})
                          \red{\lambda_j}
                          (S[u_j], C_j)
    \\[1mm]
    \textsf{(IfThenElseT)} &
                             \eval E S = \mathsf{tt} \;\Rightarrow\;
                             (S,\ifTE{E}{p}{C_1}{C_2})
                             \red{1}
                             (S, C_1)
    \\[1mm]
    \textsf{(IfThenElseF)} &
                             \eval E S = \mathsf{ff} \;\Rightarrow\;
                             (S,\ifTE{E}{p}{C_1}{C_2})
                             \red{1}
                             (S, C_2)
    \\[1mm]
    \textsf{(Call)} &
                      X \stackrel{\textcolor{sh_keyword}{\mathsf{def}}}{=} C \in \defin
                      \;\Rightarrow\;
                      (S, X) \red{{1}}^{\!\!\!\!\!\!\!\defin} (S, C)
  \end{array}
\end{displaymath}
}%
The above semantics is \emph{fully probabilistic}: for every
configuration $(S,C)$, the rules determine a unique probability
distribution over successor configurations. In particular, there is
no source of nondeterminism in the semantics. As a consequence, the
transition relation $\red{}$ induces a \emph{Discrete-Time Markov
Chain} (DTMC) when the labels $\lambda_j$ are probabilities, or a
\emph{Continuous-Time Markov Chain} (CTMC) when they are interpreted
as rates.
Intuitively, the operational rules define a global (possibly
infinite-state) Markov chain whose states are configurations of the
form $(S,C)$, and whose transitions are exactly those derivable by the
rules above, weighted by the corresponding probabilities or rates.
This global Markov chain contains all possible choreographies of the
language and all their possible states, independently of any
particular initial program.
Equivalently, one may consider the Markov chain induced by restricting
the global chain to the set of configurations reachable from a given
pair $(S,C)$. Then, the behaviour of the choreography is given by the
executions starting from this initial configuration, that is, by all
reachable (finite or infinite) paths in the Markov chain. In the
remainder, we implicitly adopt this program-indexed view.

\begin{exa}
  Consider the following choreography:
  \begin{displaymath}
    \begin{array}{lll}
      C \ {=}\ \interactBase{\nr}{\nrr}\ \lambda_1: (x'=1);\quad \interactBase{\nr}{\nrr}\
      \lambda_2: (x'=1);\quad \CEnd
    \end{array}
  \end{displaymath}
  The semantics of $C$ induces a Markov chain with states $(S_0,C)$,
  $(S_1,C')$, and $(S_1,\CEnd)$ such that
  $C'= \interactBase{\nr}{\nrr}\ \lambda_2: (x'=1);\CEnd$, $S_0(x)=0$,
  and $S_1(x)=1$. The chain can be depicted as
  \begin{center}
    \begin{tikzpicture}[ scale=0.8, transform shape, node distance=0cm
      and 4cm, conf/.style={ draw, circle, align=center,
        font=\scriptsize, inner sep=1.5pt }, ->, >=stealth ]

      \node[conf] (c0) {$
        \begin{array}{c}
          x=0\\ \hline
          C
        \end{array}$};
      
      \node[conf] (c1) [right=of c0] {$
        \begin{array}{c}
          x=1\\ \hline
          C'
        \end{array}$};
      
      \node[conf] (c2) [right=of c1] {$
        \begin{array}{c}
          x=1\\ \hline
          \CEnd
        \end{array}$};
      
      \draw[->] ([xshift=-1.2cm]c0.west) -- (c0.west);
      
      \draw (c0) -- node[above] {$\lambda_1$} (c1);
      \draw (c1) -- node[above] {$\lambda_2$} (c2);
    \end{tikzpicture}
  \end{center}
\end{exa}

\begin{exa}\label{example2}
  Consider the following choreography definition:
  \begin{displaymath}
    \begin{array}{lll}
      X \stackrel{\mathsf{def}}{=}
      \interactBase{\nr}{\nrr}
      \left\{
      \begin{array}{l}
        \lambda_1 : (x' = 1)\ \&\ (y' = 2);\ X \\[0.5mm]
        \lambda_2 : (x' = 3)\ \&\ (y' = 1);\ X
      \end{array}
      \right.
    \end{array}
  \end{displaymath}
  Let $S_0$ be a state such that $S_0(x)=0$ and $S_0(y)=0$.  Denote by
  $S_1$ and $S_2$ the states obtained by the two branches of the
  interaction, i.e., $S_1(x)=1,S_1(y)=2$ and $S_2(x)=3,S_2(y)=1$.
  The induced Markov chain for a recursive call $X$ contains six
  states, depicted as
  \begin{center}
    \begin{tikzpicture}[ scale=0.8, transform shape, node distance=0.1cm
      and 2.6cm, conf/.style={draw, circle, align=center,
        font=\scriptsize, inner sep=1.5pt}, ->, >=stealth ]

      \node[conf] (c0) {$
        \begin{array}{c}
          x=0\\ y=0\\ \hline X
        \end{array}$};
  
      \node[conf] (i0) [right=of c0] { $
        \begin{array}{c}
          x=0\\ y=0\\
          \hline
          {{\nr}\!\rightarrow\!{\nrr}\{...\}}
        \end{array}
        $};
    
      \node[conf] (c1) [above right=of i0]
      {$
        \begin{array}{c}
          x=1\\ y=2\\
          \hline
          X
        \end{array}$};
      
      \node[conf] (c2) [below right=of i0] {$
        \begin{array}{c}
          x=3\\ y=1\\ \hline X
        \end{array}$};
      
      \node[conf] (i1) [right=of c1] {$
        \begin{array}{c}
          x=1\\ y=2\\ \hline
          {{\nr}\!\rightarrow\!{\nrr}\{...\}}
        \end{array}$};
      
      \node[conf] (i2) [right=of c2] {$
        \begin{array}{c}
          x=3\\ y=1\\ \hline
          {{\nr}\!\rightarrow\!{\nrr}\{...\}}
        \end{array}$};
      
      \draw[->] ([xshift=-1.2cm]c0.west) --  (c0.west);
      
      \draw (c0) -- node[above] {$1$} (i0);
      
      \draw (i0) -- node[above left] {$\lambda_1$} (c1); \draw (i0) --
      node[below left] {$\lambda_2$} (c2);
  
      \draw (c1) -- node[above] {$1$} (i1); \draw (c2) -- node[below]
      {$1$} (i2);
      
      \draw[bend right=40] (i1) to node[above] {$\lambda_1$} (c1);
      \draw[bend right=17] (i1) to node[above] {$\lambda_2$} (c2);
      
      \draw[bend left=16] (i2) to node[below] {$\lambda_1$} (c1);
      \draw[bend left=30](i2) to node[below] {$\lambda_2$} (c2);
    \end{tikzpicture}
  \end{center}
\end{exa}

\subsection{Other language constructs}\label{sub:otherconstructs}
Our language includes constructs that are not part of the formal syntax defined above.
These constructs are purely syntactic sugar and can be easily encoded. 
Below, we discuss each of them:
\begin{itemize}

\item {\em Parametric modules.} In our implemented language, modules
  can be parameterised (indexed) as done in PRISM. We denote
  parameterised modules as $\role p[n]$ for $n$ ranging some finite
  set $N$.
  As an example, the choreography
  \[\interactBase {p[i]}{q[i]}:\lambda:U;\quad 
    \interactBase r{q[i]}:\lambda:U;\quad \CEnd
  \]
  can be easily encoded as:
  \[
    \begin{array}{llll}
      \interactBase {p1}{q1}:\lambda:U;\quad \\
      \interactBase {p2}{q2}:\lambda:U;\quad \\
      \interactBase {p3}{q3}:\lambda:U;\quad \\
      \interactBase r{q1}:\lambda:U;\quad \\
      \interactBase r{q2}:\lambda:U;\quad \\
      \interactBase r{q3}:\lambda:U;\quad \CEnd
    \end{array}
  \]
  Additionally, the choreography
  \[\interactBase {p[i]}{q[i]}:\lambda:U;\quad 
    \interactBase {q[i+1]}{p[i]}:\lambda:U;\quad \CEnd
  \]
  can be encoded in our model language as
  \[
    \begin{array}{llll}
      \interactBase {p1}{q1}:\lambda:U;\quad \\
      \interactBase {p2}{q2}:\lambda:U;\quad \\
      \interactBase {p3}{q3}:\lambda:U;\quad \\
      \interactBase {q2}{p1}:\lambda:U;\quad \\
      \interactBase {q3}{p2}:\lambda:U;\quad \\
      \interactBase {q1}{p3}:\lambda:U;\quad  \CEnd
    \end{array}
  \]
  Given that parameters range over a finite set, this operation is
  redundant. 

\item {\em The {\sf foreach} construct.}\newmark{R1.5} The use of
  parametric modules can be further facilitated by introducing
  syntactic sugar that allows iteration over the set of indices that
  parameterise these modules. To this end, our language implementation
  includes the construct
  $\textsf{foreach } (k\ \textit{op}\ i)\ u@A[k]$, which can be used
  in front of and update $u$ to parameterise it with respect to
  multiple instances of the same type of variable over different
  modules.

  \new{The construct enables the expression $(k\ \textit{op}\ i)$,
  where $k$ binds in $u@A[k]$, to identify a set of indices which
  can then be used in the update $u@A[i]$.
  Such set is obtained from the predicate determined by the operator
  $\textit{op}$ with respect to the distinguished index $i$.
  Formally, $(k\ \textit{op}\ i)$ denotes the set
  $\{\, k \in I \mid \textit{op}(k,i) \,\}$.
  For example, $(k \neq i)$ denotes
  $\{\, k \in I \mid k \neq i \,\}$, that is, all defined indices
  except~$i$.}

  In essence, the {\sf foreach} construct provides a concise and
  readable way to express operations over multiple indexed modules. It
  can be encoded explicitly by enumerating the modules manually,
  provided that the indices are known at compile time rather than
  determined dynamically at runtime.

\item {\em The {\sf \textcolor{sh_keyword}{allsynch}}
  construct.}\newmark{R1.6} Our implementation supports
  $\allsynch pGiI$, where $G$ has the form
  $\chorcommand g\lambda u$. \new{The construct requires all modules
  $\nr_i$ to synchronise while independently choosing a branch,
  unlike the interaction $\interact{\nr}{\nr_1,\ldots,\nr_n}$, where
  the chosen branch must be taken by all modules. This provides a
  structured mechanism for defining interactions in which multiple
  modules must coordinate, while their precise behaviour may vary
  dynamically (with a certain probability distribution or rate).} For
  example, consider the synchronisation between modules $\role p$ and
  $\role q$:
  \begin{displaymath}
    \allsynchName\left\{
      \begin{array}{lll}
        \nr  : (x=5) \rightarrow 42 : (x'=0) \\[1mm]
        \nr  : (x=5) \rightarrow 67 : (x'=100) \\[1mm]
        \nr  : (x=5\ \&\ z=3) \rightarrow 10 : (x'=13) \\[1mm]
        \nrr : (y=1) \rightarrow 1 : (y'=0)
      \end{array}
    \right\} ; \CEnd
  \end{displaymath}
  \new{The above choreography states that (i) $\nr$ and $\nrr$ must
  synchronise, and (ii) $\nr$ can execute any of the three declared
  updates with the respective rates, provided their guards are
  satisfied. Such behaviour partially departs from the usual {\em
    global} description given by a choreography. In this case, there
  is indeed a global synchronisation between the modules, but it is
  not specified which command must be paired with which. In a state
  where $x=5$ and $z=3$, $\nr$ can execute any of these
  actions. However, in a state where $x=5$ but $z\neq 3$, only one of
  the first two updates can be carried out.}

  A key advantage of this construct is its compactness and readability
  in specifying this type of interactions. In the scenario above,
  where modules $\role{p}$ and $\role{q}$ participate in a
  synchronised exchange, the $\allsynchName$ syntax allows for a
  concise definition of conditions under which each module updates its
  state. This avoids the need for manually encoding synchronisation
  through nested conditional constructs.
  \new{Despite its usefulness, the $\allsynchName$ construct does not
  introduce any new expressiveness but serves as syntactic sugar for
  an equivalent formulation.}  In fact, the example above can be
  rewritten by nesting $\mathtt{if\text{-}then\text{-}else}$
  constructs, ensuring that the synchronisation conditions are met:
  \begin{displaymath}
    \begin{array}{ll}
      & \mathtt{if}\ (x=5\ \&\ z=3)@\nr\ \mathtt{then}\\[1mm]
      & \quad \mathtt{if}\ (y=1)@\nrr\ \mathtt{then}\\[1mm]
      & \quad \quad \interactBase{\nr}{\nrr}
        \left\{
        \begin{array}{rlll}
          42 : &  (x'=0)\,\&\,(y'=0);\CEnd \\[1mm]
          67 : &  (x'=100)\,\&\,(y'=0);\CEnd \\[1mm]
          10 : &  (x'=13)\,\&\,(y'=0);\CEnd
        \end{array}
        \right.
      \\
      & \quad \mathtt{else}\ \CEnd \\
      & \mathtt{else}\ \mathtt{if}\ (x=5\ \&\ z\neq3)@\nr\
        \mathtt{then}\\[1mm]
      & \phantom{\mathtt{else}\ }\quad
        \mathtt{if}\ (y=1)@\nrr\ \mathtt{then}\\[1mm]
      & \phantom{\mathtt{else}\ \quad}\quad
        \interactBase{\nr}{\nrr}
        \left\{
        \begin{array}{rlll}
          42 : &  (x'=0)\,\&\,(y'=0);\CEnd \\[1mm]
          67 : &  (x'=100)\,\&\,(y'=0);\CEnd \\[1mm]
        \end{array}
        \right.
      \\
      & \phantom{\mathtt{else}\ }\quad\mathtt{else}\ \CEnd
      \\
      & \phantom{\mathtt{else}\ }\mathtt{else}\ \CEnd
    \end{array}
  \end{displaymath}
  \new{The two formulations are equivalent, up to the evaluation steps
  of the conditionals, with the $\allsynchName$
  syntax acting as syntactic sugar that provides a structured and
  readable representation of synchronisation. In general, its
  encoding into the core choreographic language requires determining
  non-overlapping guards in advance. For example, from the three
  guards $(x=5)$, $(x=5)$, and $(x=5\ \&\ z=3)$, it is necessary to
  derive the conditional guards $(x=5\ \&\ z=3)$ and
  $(x=5\ \&\ z\neq 3)$. This process depends on the functions
  $f(\tilde E)$ used in guards.}

  \new{Note that the $\allsynchName$ construct is more efficient (in
  terms of code size) than manually encoding synchronisation with
  nested $\mathtt{if\text{-}then\text{-}else}$ statements.  As the
  number of synchronising modules and conditions increases (in our
  example, there is only a single choice for $\nrr$), the depth and
  complexity of the nested conditionals grow exponentially, making the
  explicit formulation harder to read, write, and process.  In
  contrast, $\allsynchName$ provides a compact and structured way to
  express the same logic without the combinatorial explosion of nested
  conditionals.}
\end{itemize}

\section{The PRISM Language}\label{sec:prism}
We now give a formal definition of a fragment of the PRISM language by
introducing its formal syntax and semantics.

\subsection{Syntax.} We reuse some of the syntactic terms used for our
choreography language, including assignments and expressions. In the
sequel, let $a$ range over a (possibly infinite) set of labels
$\mathcal L$. We define the syntax of (a subset of) the PRISM language
as follows:
\begin{displaymath}
  \begin{array}{lrlll@{\qquad}l}
    \text{(Networks)}  \quad
    & N, M  & ::=  &      & \CEnd & \text{empty network}\\[1mm]
    &       &      & \mid & \role{p}:\{F_i\}_i & \text{module}\\[1mm]
    &       &      & \mid & M \ppp A M & \text{parallel composition}\\
    \\
    \text{(Commands)}  \quad
    & F     & ::=  &      & \command \alpha g\lambda u
                                  & \text{} (\alpha\in\{\epsilon\}\cup\mathcal L) \\
  \end{array}
\end{displaymath}
Networks are the top syntactic category for system of modules composed
together. The term $\CEnd$ represent an empty network. A module is
meant to represent a process running in the system and is denoted by
its name and its commands, formally written as $\role{p}:\{F_i\}_i$,
where $\role p$ is the name and the $F_i$'s are commands. Networks can
be composed in parallel, in a CSP style: a term like $M_1 \ppp A M_2$
says that networks $M_1$ and $M_2$ can synchronise using labels in the
finite set $A$.
In this work, we omit PRISM's hiding and substitution constructs as
they are irrelevant for our current choreography language.
Commands in a module have the form
$[\alpha] g \rightarrow \Sigma_{i\in I}\{\lambda_i: u_i\}$. The
character $\alpha$ can either be the empty string $\epsilon$ or a
label $a$, i.e., $\alpha\in\{\epsilon\}\cup \mathcal L$. If $\epsilon$
then no synchronisation is required. On the other hand, if there is
label $a$ then there will be a synchronisation with other modules that
must synchronise on $a$. The term $g$ is a guard on the current
variable state. If both label and guard are enabled, then the command
executes a branch $i$ with probability/rate $\lambda_i$. As for
choreographies, if the $\lambda_i$'s are probabilities, we must have
that $0\leq\lambda_i\leq 1$ and $\Sigma_{i\in I}\lambda_i=1$.

\subsection{Semantics.} To give a probabilistic semantics to the PRISM
language, we follow the approach given in the PRISM
documentation~\cite{PRISMdoc}.  Hereby, we do that by defining two
relations: one with labels for networks and one on states. Our
relation on networks is the smallest relation $\prismred{}{}$
satisfying the rules given in Fig.~\ref{fig:semantics}.
\begin{figure}[h]
  {\small
  \begin{displaymath}
    \begin{array}{ccc}
      \infer[\mathsf{(M)}]
      { {\role{p}:\{F_k\}_k} \prismred{} F
      }
      {
      F\in\{F_k\}_k
      }
      \quad
      \infer[\mathsf{(P_1)}]
      {M_1\ppp A M_2\prismred{}\command {\alpha}g\lambda u
      }
      {
      \exists j\in\{1,\!2\}.\ 
      M_j\prismred{} \command {\alpha}g\lambda u
      & \alpha\!\not\in\! A
        }
      \\\\
      \infer[\mathsf{(P_2)}]
      {M_1\ppp A M_2\prismred{}
      \commandBase a{g\land g'}
      {\Sigma_{i,j}\,\lambda_i*\lambda'_j}: {u_i\& u'_j} 
      }
      {
      M_1\prismred{}\command {a}g\lambda u
      \quad&\quad 
        M_2\prismred{}\commandBase {a} {g'} \Sigma_{j\in J}\lambda'_j: u'_j
      \ &\
             a\!\in\! A}
    \end{array}
  \end{displaymath}
  }%
  \caption{Semantics for PRISM networks} 
  \label{fig:semantics}
\end{figure}
Rule $\mathsf{(M)}$ just exposes a command at network level. Rule
$\mathsf{(P_1)}$ propagates a command through parallel composition if
$\alpha$ is empty or if the label $a$ is not part of the set $A$. When
the label $a$ is in $A$, we apply rule $\mathsf{(P_2)}$. In this case,
the product of the probabilities/rates must be taken by extending the
two different branches to every possible combination. This also
includes the combination of the associated assignments.

Based on the relation above, given
$M\prismred{}\command \alpha g\lambda u$ and two
states $S$ and $S'$, we define the function
$$\mu(\command \alpha g\lambda u,\ S,\ S'\ ) = \Sigma_{S[u_i]=S',i\in I}\lambda_i$$
which gives the probability/rate for the system to go from state $S$
to state $S'$ after executing command $\command \alpha g \lambda u$,
for some $\alpha$.
If the $\lambda_i$ are probabilities, then the function must be a
probability distribution. Note that $\mu(F,S,S')$ only denotes the
probability/rate for the system to move from state $S$ to state $S'$
after executing command $F$. However, there can be other commands
derived from a given network $M$ through the relation $\prismred{}$
that would cause a transition from $S$ to $S'$. Therefore, we define
the transition relation on states $M\vdash S\red{\lambda}S'$ as
$$
\infer[\mathsf{(Transition)}]
{
  M\vdash S\red{\sum_{j} \mu(F_j,S,S')}S'
}
{
  \forall j,\alpha.\ M\prismred{}F_j
  \quad&\quad
  S\vdash F_j
}
$$
where $S\vdash \command \alpha g\lambda u$ is defined as $\eval gS$.
Note that since PRISM is declarative, a term $M$ never changes while
the state of the system evolves.

It is important to point out that, in general, the transition rule
above does not give the exact probability of a transition in case of a
Markov chain (DTMC), since the sum ${\sum_{j} \mu(F_j,S,S')}$ could be
a value greater than $1$. In order to get the right probability, the
value has to be normalised for all reachable $S'$. In the next
section, we will show that this is not an issue for networks that are
obtained from our translation from choreography to PRISM.

\begin{exa}
  Consider the following network $M$:
  \begin{displaymath}
    \begin{array}{llll}
      \role{p}: \{ & \quad \commandBase {}{x=0} {1:(x'=1)}\\
                   & \quad \commandBase {a}{y<1} {0.4: (x'=x+1)\ +\ 0.6: (x'=x)}\quad \}\\[2mm]
      \role{q}: \{ & \quad \commandBase {}{y=0} {1:(y'=1)}\\
                   & \quad \commandBase {a}{x<1} {0.5: (y'=y+1)\ +\ 0.5: (y'=y)}\quad \}\\
    \end{array}
  \end{displaymath}
  above, the two modules $\role p$ and $\role q$ can both do
  independent actions, as well as synchronising on label $a$. Applying
  the semantics, we can easily derive
  $M\prismred{}\commandBase {}{x=0} {1:(x'=1)}$,
  $M\prismred{}\commandBase {}{y=0} {1:(y'=1)}$, and $M\prismred{}F$,
  such that
  {\small
  \begin{displaymath}
    \begin{array}{rlll}
      F=\commandBase {a}{x<1\ \&\ y<1} {\quad} & 0.2: (x'=x+1)\ \&\ (y'=y+1)\\
                                       + & 0.2: (x'=x+1)\ \&\ (y'=y)\\
                                       + & 0.3: (x'=x)\ \&\ (y'=y+1)\\
                                       + & 0.3: (x'=x)\ \&\ (y'=y)\\
    \end{array}
  \end{displaymath}
  }%
  Let $S_0$, $S_1$, $S_2$, and $S_3$ be all possible reachable states
  such that
  $S_0(x)=S_0(y)=0$, $S_1(x)=1$, $S_1(y)=0$, $S_2(x)=0$, $S_2(y)=1$,
  and $S_3(x)=S_3(y)=1$.
  Moreover, let $S_0$ be the starting state. Then, 
  \begin{displaymath}
    \begin{array}{ll}
      \mu(\commandBase {}{x=0} {1:(x'=1)}, S_0, S_1) = 1
      & \quad \mu(\commandBase {}{x=0} {1:(x'=1)}, S_2, S_3) = 1\\
      \mu(\commandBase {}{y=0} {1:(y'=1)}, S_0, S_2) = 1
      & \quad \mu(\commandBase {}{y=0} {1:(y'=1)}, S_1, S_3) = 1\\ 
      \mu(F, S_0, S_1) = 0.2 &\quad\mu(F, S_0, S_2) = 0.3\\
      \mu(F, S_0, S_0) = 0.3 & \quad\mu(F, S_0, S_3) = 0.2\\
    \end{array}
  \end{displaymath}  
  Now, by rule $\textsf{(Transition)}$ above, we have that
  $M\vdash S_0\rightarrow_{1.2} S_1$,
  $M\vdash S_0\rightarrow_{1.3} S_2$,
  $M\vdash S_0\rightarrow_{0.2} S_3$, and
  $M\vdash S_0\rightarrow_{0.3} S_0$. Clearly, both transitions should
  be normalised, finally yielding the following DTMC:
  \begin{center}
    \begin{tikzpicture}[
      scale=0.8,
      transform shape,
      node distance=0.4cm and 3cm,
      conf/.style={
        draw, circle,
        align=center,
        font=\scriptsize,
        inner sep=1.5pt
      },
      ->, >=stealth
      ]

      \node[conf] (S0) {$
        \begin{array}{c}
          S_0
        \end{array}$};

      \node[conf] (S1) [above right=of S0] {$
        \begin{array}{c}
          S_1
        \end{array}$};

      \node[conf] (S2) [below right=of S0] {$
        \begin{array}{c}
          S_2
        \end{array}$};
      
      \node[conf] (S3) [below right=of S1] {$
        \begin{array}{c}
          S_3
        \end{array}$};

      \draw (S0) edge[loop above] node{$0.1$} (S0);
      \draw[bend left=30](S0) to node[above] {$0.4$} (S1);
      \draw[bend right=30] (S0) to node[below] {$0.43$} (S2);
      \draw (S0) to node[above right] {$0.07$} (S3);
      
      \draw[bend left=30] (S1) to node[above] {$1$} (S3);
      \draw[bend right=30] (S2) to node[below] {$1$} (S3);
      
    \end{tikzpicture}
  \end{center}
\end{exa}

\begin{exa}\label{example3}
  The choreography presented in Example \ref{example2} can be
  described by the following PRISM network $M$ (for
  $\lambda_i=\mu_i*\gamma_i$):
{\small
\begin{displaymath}
  \begin{array}{llll}
    \role{p}: \{ & \quad \commandBase {a}{s_{\role p}=0} {\mu_1:(x'=1)
                   \ \&\ (s_{\role p}'=1)} 
                   \quad\qquad \commandBase {}{s_{\role p}=1} 
                   {1:(s_{\role p}'=0) }\\
                 & \quad \commandBase {b}{s_{\role p}=0} {\mu_2:(x'=3)
                   \ \&\ (s_{\role p}'=2)}
                   \quad\qquad \commandBase {}{s_{\role p}=2} 
                   {1:(s_{\role p}'=0) }\quad \}\\[2mm]
    \role{q}: \{ & \quad \commandBase {a}{s_{\role q}=0} {\gamma_1:(y'=2)
                   \ \&\ (s_{\role q}'=1) } 
                   \qquad\quad \commandBase {}{s_{\role q}=1} 
                   {1:(s_{\role q}'=0) }\\
                 & \quad \commandBase {b}{s_{\role q}=0} {\gamma_2:(y'=1)
                   \ \&\ (s_{\role q}'=2) } 
                   \qquad\quad \commandBase {}{s_{\role q}=2} 
                   {1:(s_{\role q}'=0) }\quad \}\\
  \end{array}
\end{displaymath}
}%
The two modules $\role p$ and $\role q$ synchronise on the labels $a$
and $b$.  Applying the semantics, the global state can evolve
according to $F_1$ or $F_2$, defined as follows:
{\small
\begin{displaymath}
  \begin{array}{rlll}
    F_1=\commandBase {a}{s_{\role p}=0\ \&\ s_{\role q}=0} {\quad} \mu_1*\gamma_1 : 
    (x'=1) \ \&\ (y'=2) \ \&\ (s_{\role p}'=1) \ \&\ (s_{\role q}'=1) \\
    F_2=\commandBase {b}{s_{\role p}=0\ \&\ s_{\role q}=0} {\quad} \mu_2*\gamma_2 : 
    (x'=3) \ \&\ (y'=1) \ \&\  (s_{\role p}'=2) \ \&\ (s_{\role q}'=2)\\

  \end{array}
\end{displaymath}
}

\end{exa}

\section{Projection}\label{sec:proj}
In this section, we provide a treatment of projection, which
constitutes the mapping from choreographies to the PRISM language.

\subsection{Mapping Choreographies to PRISM} 
The process of generating endpoint code from a choreography is
commonly referred to as \emph{projection}. Typically, projection is
defined separately for each module appearing in the choreography
program: given a module (often called a role) and a choreography, it
generates the code for that particular module.
In our approach, projection targets PRISM modules and is realised
using the mechanisms provided by the PRISM modelling language. In
particular, a choreography interaction is simulated in PRISM by using
(i) labels on which the involved modules synchronise and (ii) state
variables that enable the appropriate commands at the correct points
in the execution.

Before formalising this idea, we introduce additional machinery that
allows us to define projection. To this end, we annotate choreographies
with labels. We refer to such choreographies as \emph{annotated
choreographies}. 
We assume two disjoint classes of labels. The first class consists of
\emph{control labels}, ranged over by $a,b$, which are taken from the
PRISM set of labels $\mathcal L$. These labels will later be used as
synchronisation labels in the projection to PRISM modules. The second
class consists of \emph{state labels}, ranged over by $\iota$, drawn
from some countably infinite set, subset of all possible values $v$ in
the syntactic category for expressions $E$. State labels represent the
local control states associated with modules during the execution of
the projection of a choreography.
\newcommand{\anno}[1]{{#1}^{\mathsf a}}
\begin{defi}[Annotated Choreography]
  An \emph{annotated choreography} is obtained from the choreography
  syntax by decorating each construct with a control label and each
  module occurrence with a state label:
  {\small
  \[
    \anno{C} ::= 
      \interactl{\nr^{\iota}}a{\nr_1^{\iota_1},\ldots,\nr_n^{\iota_n}}
      \ \mid\ 
      \ifTEl{E}a{\nr^{\iota}}{\anno{C}_1}{\anno{C}_2}
      \ \mid\ 
      X^{a,\sigma}
      \ \mid\ 
      \CEnd^{\sigma}
  \]
  }
\end{defi}
\noindent
Above, $\sigma$ is an injective function assigning a state label to
each module.  The intuition behind annotations is that they allow us
to identify a particular point in the choreography, enabling the
projection of all modules involved to synchronise accordingly.
Fresh control labels are generated by $\mathsf{fresh}_{\mathsf c}$ and
fresh state labels by $\mathsf{fresh}_{\mathsf s}$. Then, we can
define the function $\mathsf{annot}$ for annotating a choreography as
follows:
\begin{defi}[Annotation]\label{def:annot}
  The annotation function $\mathsf{annot}(C,\sigma)=C^a$ is defined as
  follows:
  {\small
  \[
    \begin{array}{@{}c@{}}
      \infer
      {
      \mathsf{annot}\!\left(
      \interact{\nr}{\nr_1,\ldots,\nr_n},\ \sigma
      \right)
      \ \ =\ \
      \interactl{\nr^{\sigma(\nr)}}a{\nr_1^{\sigma(\nr_1)},\ldots,\nr_n^{\sigma(\nr_n)}}
      }
      {
      a=\mathsf{fresh}_{\mathsf c}
      \quad
      \forall j\in J.\ \mathsf{annot}(C_j,\ \sigma[
      \nr\mapsto\mathsf{fresh}_{\mathsf s},
      \nr_1\mapsto\mathsf{fresh}_{\mathsf s},\ldots,
      \nr_n\mapsto\mathsf{fresh}_{\mathsf s}])=\anno {C}_j
      }
      \\[6mm]
      \infer
      {
      \mathsf{annot}(\ifTE{E}{\nr}{C_1}{C_2},\sigma)
      =
      \ifTEl{E}a{\nr^{\sigma(\nr)}}{\anno {C}_1}{\anno C_2}
      }
      {
      a=\mathsf{fresh}_{\mathsf c}
      \quad
      \forall i\in\{1,2\}.\,\sigma_i=\sigma[\nr\mapsto\mathsf{fresh}_{\mathsf s}]
      \quad
      \mathsf{annot}(C_1,\sigma_1)=\anno C_1
      \quad
      \mathsf{annot}(C_2,\sigma_2)=\anno C_2
      }
      \\[6mm]
      \infer
      {
      \mathsf{annot}(X,\sigma)=X^{a,\sigma}
      }
      {
      a=\mathsf{fresh}_{\mathsf c}
      }
      \qquad\qquad
      \infer
      {
      \mathsf{annot}(\CEnd,\sigma)=\CEnd^{\sigma}
      }
      {
      }
\end{array}
\]
  }
\end{defi}
\noindent
The function above places a fresh control label $a$ in every language
construct (except from $\CEnd$). Additionally, it marks each
occurrence of a module $\nr$ with the state label $\sigma(\nr)$. After
a state label is used, the function updates the state assignment by
replacing it with a fresh state labels in the continuations. Note that
this mechanism does not make state labels globally unique, since a
module may be assigned the same state label on different branches
during the annotation process.
For example, consider the choreography
$
\interactBase{\nr}{\nrr} :
\{\lambda_1:u_1;\ \interactBase{\nrrr}{\nr}: \lambda_3:u_3,\quad
\lambda_2:u_2;\ \interactBase{\nrrr}{\nr}: \lambda_4:u_4\}
$.
After annotating the first interaction, modules $\nr$ and $\nrr$
receive different fresh state labels in the two
continuations. However, module $\nrrr$ is not involved in the first
interaction and therefore retains the same state label $\sigma(\nrrr)$
in both branches, resulting in
$
\interactBasel{\nr^{\iota_1}}a{\nrr^{\iota_2}} :
\{\lambda_1:u_1;\ \interactBasel{\nrrr^{\iota_3}}b{\nr^{\iota_4}}: \lambda_3:u_3,\quad
\lambda_2:u_2;\ \interactBasel{\nrrr^{\iota_3}}c{\nr^{\iota_5}}: \lambda_4:u_4\}
$.
Each definition
$X\stackrel{\textcolor{sh_keyword}{\mathsf{def}}}{=} C$ in $\defin$
must be annotated as $\mathsf{annot}(C, \sigma)$ starting from a fresh
$\sigma$.

\newcommand{\toprole}{\mathsf{top}}

Given an annotated choreography, the partial function $\toprole$
returns the first state label associated with a given module in the
choreography.  Formally, it is defined as:
\[
  \begin{array}{lll}

    \toprole_{\nr}\ ( \interactl{\nr^{\iota}}a{\nr_1^{\iota_1},\ldots,\nr_n^{\iota_n}})
    = \iota
    \\[2mm]

    \toprole_{\nr_i}\ ( \interactl{\nr^{\iota}}a{\nr_1^{\iota_1},\ldots,\nr_n^{\iota_n}})
    = \iota_i
    &    \qquad (1 \le i \le n)
    \\[2mm]

    \toprole_{\nrr}\ ( \interactl{\nr^{\iota}}a{\nr_1^{\iota_1},\ldots,\nr_n^{\iota_n}})
    = \toprole_{\nrr}(\anno C_1)
    &  \quad (\nrr\not\in\{\nr,\nr_1,\ldots,\nr_n\},
    \\
    &  \quad\ \forall i,j.\
      \toprole_{\nrr}(\anno C_i)=\toprole_{\nrr}(\anno C_j))
    \\[2mm]

    \toprole_{\nr}\ (\ifTEl{E}a{\nr^{\iota}}{\anno C_1}{\anno C_2})
    = \iota
    \\[2mm]

    \toprole_{\nrr}\ ( \ifTEl{E}a{\nr^{\iota}}{\anno C_1}{\anno C_2})
    = \toprole_{\nrr}(\anno C_1)
    & \quad (\nrr\neq\nr,
    \\
    & \quad\ \toprole_{\nrr}(\anno C_1)=\toprole_{\nrr}(\anno C_2))
    \\[2mm]

    \toprole_{\nr}\ ( X^{a,\sigma})
    = \sigma(\nr)
    \qquad\qquad
    \toprole_{\nr}\ (\CEnd^{\sigma})
    = \sigma(\nr)
\end{array}
\]

\noindent
We observe that $\toprole$ is always defined if the annotation is
obtained from $\mathsf{annot}$. More precisely, given a choreography
$C$, a module $\nr$, and a $\sigma$, we have that there exists $\iota$
such that $\toprole_{\nr}(\mathsf{annot}(C,\sigma))=\iota$.

Given a label $a$ and a set of indices $I$, the set $\{a_i|i\in I\}$
contains unique identifiers generated from $a$.  We do similarly for
state labels $\iota$. Moreover, we extend a state with special
variables $s_{\nr}$, one for each module that we are considering in a
given choreography.
The intuition is to use this variable to store the state label of the
choreography that its projection is simulating. This is in line with
the example in Section~\ref{sec:intro}, except that our formal
projection uses annotation labels instead of natural numbers.
We can then define the projection function. Since there are key
differences between using probabilities and using rates, we proceed
separately. We begin with choreographies that involve rates:
\begin{defi}[Projection, CTMC]\label{def:projCTMC} Given
  an annotated choreography with rates $\anno C$ and a module
  $\nrr$, 
  we define the function $\proj$ as:
  {\small
  \begin{displaymath}
    \begin{array}{@{}lr@{}}
      \proj
      (\nrr,\interactl{\nr^{\iota}}{a}{\nr_1^{\iota_1},\ldots,\nr_n^{\iota_n}})=
      &  \boxed{\text{if }\nrr=\nr}\\[2mm]
      \quad
      \left\{\commandBase {a_j} {\ s_{\nrr}\!=\! \iota} {\lambda_j:}
      \ s_{\nrr}'\!=\! \mathsf{top}_{\nrr}(\anno{C}_j)\ 
      \&\ \projE {u_j}\nrr\right\}_{j\in J}
      \,\cup\, \bigcup_{j\in J} \proj (\nrr, \anno{C}_j)
      \\[1cm]
          
      \proj
      (\nrr,\interactl{\nr^{\iota}}{a}{\nr_1^{\iota_1},\ldots,\nr_n^{\iota_n}})=
      &  \boxed{\text{if }
        \nrr= \nr_i}\\[2mm]
      \quad
      \left\{\commandBase {a_j} {\ s_{\nrr}\!=\! \iota_i} {1:\ s_{\nrr}'\!=\!
      \mathsf{top}_{\nrr}(\anno{C}_j)
      }
      \ \&\ \projE {u_j}\nrr\right\}_{j\in J}
      \,\cup\, \bigcup_{j\in J} \proj (\role q, \anno{C}_j)
      \\[1cm]
      
      \proj
      (\nrr,\interactl{\nr^\iota}a{\nr_1^{\iota_1},\ldots,\nr_n^{\iota_n}})\ =
      \ \bigcup_{j\in J} \proj (\role q, \anno{C}_j)
      &  \boxed{\text{if }\role q\not\in\{\role p, 
        \nr_1,\ldots,\nr_n\}}
      \\[1cm]

      \proj (\nrr,\ifTEl {E}a{\nr^\iota}{\anno C_1}{\anno C_2}) = 
      &  \boxed{\text{if }\nrr=\nr}\\[2mm]
      \quad\left\{ 
      \begin{array}{lll}
        \commandBase {} {s_{\nrr}\!=\! \iota\ \&\ E}{\ 1: s'_{\nrr}\!=\!
        \mathsf{top}_{\nrr}(\anno C_1)},\\ 
        \commandBase {} {s_{\nrr}\!=\! \iota\ \&\ \mathsf{not}(E)}
        {\ 1: s'_{\nrr}\!=\! \mathsf{top}_{\nrr}(\anno C_2)}
      \end{array}
      \right\}
      \ \cup\ \proj (\nrr, \anno C_1)
      \ \cup\
      \proj (\nrr, \anno C_2)
      \\[1cm]

      \proj (\nrr,\ifTEl {E}a{\nr^\iota}{\anno C_1}{\anno C_2}) \ =\
      \proj (\nrr, \anno C_1)
      \ \cup\
      \proj (\nrr, \anno C_2)
      &  \boxed{\text{if }\nrr\neq\nr}
      \\[1cm]

      \proj (\nrr,\CEnd^{\sigma}) = \emptyset
      \quad
      \proj (\nrr, X^{a,\sigma}) =
      \{ \commandBase {a} {s_{\nrr}\!=\! \sigma(\nrr)}{\ 1: s'_{\nrr}\!=\! } \mathsf{top}_{\nrr}(\anno C)\}
      &\boxed{X\stackrel{\textcolor{sh_keyword}{\mathsf{def}}}{=}
        \anno C\in\defin}
    \end{array}
  \end{displaymath}
  }
\end{defi}
\noindent
We examine the various cases in the definition above. Intuitively, the
projection function makes sense only for choreographies that have been
annotated using $\mathsf{annot}$. The first three cases deal with the
projection of an interaction.
When projecting the first module $\nr$, we create one command
$\commandBase {a_j} {s_{\nrr}\!=\! \iota} {\lambda_j:\ s_{\role
    q}'\!=\!  \topannotation \nrr{\anno C_j}}\ \&\ \projE {u_j}{\nrr}$
for each branch such that
\begin{itemize}
\item the label $a_j$ and its uniqueness ensure that all modules take
  the same branch;
\item the guard $s_{\role q}\!=\! \iota$ ensures that these commands
  are only executable when the system has reached this particular
  state, identified by the reserved variable $s_{\role q}$;
\item the rate $\lambda_j$ is the rate that appears in the same branch
  of the choreography;
\item the successor state is determined by updating $s_{\role q}$ to
  $\topannotation \nrr{C_j}$, depending on which branch $j$ was
  selected. 
  The function $\mathsf{annot}$ ensures that every step in all
  branches is assigned to a different label, thereby also discarding
  all branches that are not selected;
  
\item the projected update $\projE{u_j}{\nrr}$ acts as a filter on the
  list of updates in $u_j$, ensuring that only those variables local
  to $\role q$ are updated. 
  We do not define $\projE{u_j}{\nrr}$ explicitly, but we assume that
  it adheres to the allocation of variables fixed prior to projection.
  Since PRISM would reject an input in which multiple modules update
  the same variables, our projection is undefined in such cases.

\end{itemize}

\noindent
The second case defines the projection of an interaction for one of
the modules \( \{\role p_1, \ldots, \role p_n\} \). Similarly to the
previous case, we define a command for each branch of the
interaction. However, the rate of each command is set to~1, ensuring
that each branch synchronises with rate \( \lambda_j \cdot 1 \) (see
rule \( \mathsf{(P_2)} \) in Figure~\ref{fig:semantics}).
The third case concerns the projection of a module that does not
belong to the set \( \{\nr, \nr_1,\ldots,\nr_n\} \). In this case, the
projection collects the commands generated from all $\anno C_j$. Note
that $\mathsf{annot}$ assigns the same state label: this is not a
problem, since any synchronisation is performed on different control
labels.
The if--then--else construct focuses on the module \( \nr \), where
the guard \(E\) is evaluated. Being this a local operation, no
synchronisation labels are required. For recursive calls, we generate
a command that redirects the execution to the projection of the
definitions \(\defin\), which are projected separately.
Formally, the commands of a module $\nrr$ are those generated
from the choreography together with those generated from the body of
every definition in $\defin$:
\[
  \projD(\nrr,\anno C)\ =\ \proj(\nrr,\anno C)\ \cup
  \bigcup_{(X \defrec \anno C_X)\in\defin} \proj(\nrr,\anno C_X)
\]
The \emph{projected network} of $\anno C$ is then
$\mathcal N(\anno C) = \pp_{\nrr}\ \nrr:\projD(\nrr,\anno C)$, where
$\nrr$ ranges over the modules occurring in $\anno C$ or in $\defin$,
and $\pp$ denotes alphabetised parallel composition, in which modules
synchronise only on shared labels.

\begin{exa}\label{example-proj}
  In order to illustrate how our projection works, consider the
  following example, in which we apply the projection to the
  choreography $X$ from Example~\ref{example2}. This yields PRISM
  modules that are slightly different from those presented in
  Example~\ref{example3}, but which implement a similar behaviour.
  In Example \ref{example2}, we defined a recursive choreography in
  which module \(\nr\) interacts with module \(\nrr\) through two
  branches. Its annotated form can be written as:
  \begin{displaymath}
    \begin{array}{lll}
      X \defrec \interactBasel{\nr^{\iota_1}}a{\nrr^{\iota_2}}
      \left\{
      \begin{array}{lll}
        \lambda_1: (x'=1)\&(y'=2);\ X^{b,\sigma_1}
        \\
        \lambda_2: (x'=3)\&(y'=1);\ X^{c,\sigma_2}
      \end{array}
      \right.
    \end{array}
  \end{displaymath}
  where
  $\sigma_i=\sigma[\nr\mapsto\mathsf{fresh}_{\mathsf
    s},\nrr\mapsto\mathsf{fresh}_{\mathsf s}]$ for $i\in\{1,2\}$ and
  for some $\sigma$ such
  that $\sigma(\nr)=\iota_1$ and $\sigma(\nrr)=\iota_2$. Our
  projection, as defined in Definition~\ref{def:projCTMC}, when
  applied to a given choreography $X^{d,\sigma'}$ (which essentially
  corresponds to a recursive call), generates the following modules:
  {\small
  \begin{equation}\label{eq:1}
    \begin{array}{llll}
      \role{p}: \left\{
      \begin{array}{lll}
        \commandBase {d}{s_{\nr}=\sigma'(\nr)} 
        {1:(s_{\nr}'=\iota_1) },
        \\
        \commandBase {a_1}{s_{\nr}=\iota_1} {\lambda_1:(x'=1)
        \ \&\ (s_{\nr}'=\sigma_1(\nr))},
        & \quad \commandBase {b}{s_{\nr}=\sigma_1(\nr)} 
          {1:(s_{\nr}'=\iota_1) },\\
        \commandBase {a_2}{s_{\nr}=\iota_1} {\lambda_2:(x'=3)
        \ \&\ (s_{\nr}'=\sigma_2(\nr))},
        & \quad \commandBase {c}{s_{\role p}=\sigma_2(\nr)} 
          {1:(s_{\role p}'=\iota_1) }
      \end{array}
      \right\}
      \\[7mm]
      \nrr:
      \left\{
      \begin{array}{lll}
        \commandBase {d}{s_{\nrr}=\sigma'(\nrr)} 
        {1:(s_{\nrr}'=\iota_2) },
        \\
        \commandBase {a_1}{s_{\role q}=\iota_2} {1:(y'=2)
        \ \&\ (s_{\role q}'=\sigma_1(\nrr)) }, 
        & \quad \commandBase {b}{s_{\role q}=\sigma_1(\nrr)} 
          {1:(s_{\role q}'=\iota_2) }, \\
        \commandBase {a_2}{s_{\role q}=\iota_2} {1:(y'=1)
        \ \&\ (s_{\role q}'=\sigma_2(\nrr)) },
        & \quad \commandBase {c}{s_{\role q}=\sigma_2(\nrr)} 
          {1:(s_{\nrr}'=\iota_2) }
      \end{array}
      \right\}
    \end{array}
  \end{equation}
  }%
  The core idea is that we use reserved variables $s_{\nr}$ and
  $s_{\nrr}$ to track the state of each module, in order to simulate
  the behaviour of the choreography.
  The various steps of the projection can then be summarised as
  follows:
  \begin{enumerate}
  \item We first project \(X^{d,\sigma'}\). This generates the
    commands
    \(\commandBase {d}{s_{\nr}=\sigma'(\nr)} {1:(s_{\nr}'=\iota_1) }\)
    and
    \(\commandBase {d}{s_{\nrr}=\sigma'(\nrr)} {1:(s_{\nrr}'=\iota_2)
    }\) for \(\nr\) and \(\nrr\), respectively.  Since
    \(X^{d,\sigma'}\) is the starting choreography, we assume that all
    PRISM modules initially start in a state where the reserved
    variables satisfy \(s_{\nr}=\sigma'(\nr)\) and
    \(s_{\nrr}=\sigma'(\nrr)\).  The intention is that the commands of
    the two modules synchronise on \(d\) and then activate the first
    action in the body of \(X\).
    
  \item Separately, we must project the bodies of all definitions in
    \(\defin\), only one in this case.
    The first step of the projection ensures that this interaction is
    enabled only when the variables \(s_{\nr}\) and \(s_{\nrr}\) are
    equal to the label $\iota_1$ and $\iota_2$ respectively. The
    operation has two possible outcomes: either a synchronisation on
    label \(a_1\) with rate \(\lambda_1\), or a synchronisation on
    label \(a_2\) with rate \(\lambda_2\).  In the first case, both
    \(s_{\nr}\) and \(s_{\nrr}\) are set to $\sigma_1(\nr)$ and
    $\sigma_1(\nrr)$, respectively.  In the second case, \(s_{\nr}\)
    and \(s_{\nrr}\) are respectively set to $\sigma_2(\nr)$ and
    $\sigma_2(\nrr)$.

  \end{enumerate}

\end{exa}

\noindent
The projection from Definition~\ref{def:projCTMC} works only with
rates. In the case of probabilities, we must ensure that the
probabilities of a branching sum to~1, which is not guaranteed by
Definition~\ref{def:projCTMC}, since it generates, for each branch
$j$, a separate command whose single update carries probability
$\lambda_j$. Using instead a single command carrying the whole
distribution would require a single synchronisation label, so that we
could not enforce both $\nr$ and $\{\nr_1,\ldots,\nr_n\}$ to take the
same branch according to the probability distribution given by the
\(\lambda_i\)’s. To address this issue, we introduce the following
definition instead:
\begin{defi}[Projection, DTMC]\label{def:projDTMC} Given
  an annotated choreography with probabilities $\anno C$ and a module
  $\nrr$, we define $\proj$ as:
  {\small     
  \begin{displaymath}
    \begin{array}{lr}
      \proj
      (\nrr,\interactl{\nr^\iota}a{\nr_1^{\iota_1'},\ldots,\nr_n^{\iota_n'}})= 
      &  \boxed{\text{if }\nrr=\nr}\\[2mm]
      \
      \begin{array}{lll}
        \left\{\commandBase {} {s_{\nrr}\!=\! \iota}{\ \sum_{j\in J}
        \lambda_j: s'_{\nrr}\!=\! \iota_j}\right\}\quad\cup\\[2mm]
        \left\{\commandBase {a_j} {\ s_{\nrr}\!=\! \iota_j} {1:}
        \ s_{\nrr}'\!=\! \mathsf{top}_{\nrr}(\anno C_j)\
        \&\ \projE {u_j}\nrr\right\}_{j\in J}
        \ \ \cup\ \ \bigcup_{j\in J} \proj (\nrr, \anno C_j)
      \end{array}
    \end{array}
  \end{displaymath}
  }%
  The other cases of the definition are as in
  Definition~\ref{def:projCTMC}, except that in the case $\nrr=\nr_i$
  the guard is $s_{\nrr}=\iota_i'$.
\end{defi}
\noindent
The main intuition behind the above definition is that module $\nr$
takes a (probabilistic) internal decision on the $j^{\text{th}}$
branch and then synchronises on label $a_j$ with
$\{\nr_1,\ldots,\nr_n\}$.

\smallskip

\subsection{Correctness.}
Our projection guarantees a correspondence between the semantics of a
choreography and that of its projection. Below, we assume to work with
annotated choreographies. However, to make notation lighter, we do not
use annotation when not necessary.
In order to state our main theorem, we rely on the notions of
\emph{head modules} and \emph{strongly connected} choreographies. The
former identifies the modules involved in the next action of a
choreography:
\begin{defi}[Head Modules]
  The function $\hmodules$ is defined as follows:
  {\small
  \begin{displaymath}
    \begin{array}{rll}
      \hmodules(\interact{\nr}{\nr_1,\ldots,\nr_n})\
      & = \ \{\nr, \nr_1,\dots,\nr_n\}\\[1mm]
      \hmodules(\ifTE {E}{\nr}{C_1}{C_2}) \ & =\ \{\nr\}\\
      \hmodules(X) \ & =\ \hmodules (C)\qquad\qquad(\text{if }
                       X\defrec C\in\defin)\\[1mm]
      \hmodules(\CEnd) \ & =\ \emptyset
    \end{array}
  \end{displaymath}
  }
\end{defi}
\noindent
Then, the property of strongly connected is defined below.
\begin{defi}[Strongly Connected Choreography]
  A choreography $C$ is \emph{strongly connected}, written
  $\sconnected(C)$, if it satisfies the following conditions (the
  rules are read coinductively):
  {\small
  \begin{displaymath}
    \begin{array}{c}
      \infer {
      \sconnected(\interact{\nr}{\nr_1,\ldots,\nr_n})
      } {
      \forall j\in J.\ \sconnected(C_j)\ \land\ \big(\hmodules(C_j)\neq\emptyset\Rightarrow
      \hmodules(C_j)\cap\{\nr, \nr_1,\dots,\nr_n\}\neq\emptyset\big)
      }
      \qquad\qquad
      \infer {
      \sconnected(\CEnd)
      } {
      }
      \\\\

      \infer {
      \sconnected(\ifTE {E}{\nr}{C_1}{C_2})
      } {
      \forall j\in \{1,2\}.\ \sconnected(C_j)\ \land\ \big(\hmodules(C_j)
      \neq\emptyset\Rightarrow \nr\in\hmodules(C_j)\big)
      }
      \qquad\qquad
      \infer {
      \sconnected(X)
      } {
      \sconnected(C)\qquad X\defrec C\in\defin
        }
      \\\\
    \end{array}
  \end{displaymath}
  }
\end{defi}
\noindent
The notion of connectedness is quite well-known in the
literature. Since our framework is based on synchronous communication,
we follow the same approach as that of Carbone et
al.~\cite{CHY12}. The basic idea is that each interaction shares at
least one module with the subsequent choreography. In particular, in
every branch of a probabilistic choice or an if-then-else involving
modules $\nr, \nr_1, \ldots, \nr_n$, the next action (if any) must
involve one of $\nr, \nr_1, \ldots, \nr_n$, possibly after unfolding
recursive calls. Connectedness guarantees a form of causality between
interactions: subsequent interactions between modules always depend on
previous ones.
As an example of strongly connected choreography, the choreography
$X$, obtained from the definition
{\small
\begin{displaymath}
  X\defrec
  \interactBase{\nr}{\nrr}:\,
  \left(
    \begin{array}{l}
      \lambda_1: u_1;\ \interactBase{\nrr}{\nrrr}:\, \lambda_1':u_1';\ X    \\
      \lambda_2: u_2;\ X
    \end{array}
  \right)
\end{displaymath}
}%
is strongly connected while 
$
\interactBase{\nr}{\nrr}:\,
\left(
  \begin{array}{l}
    \lambda_1: u_1;\ \interactBase{\nrrr_1}{\nrrr_2}:\, \lambda_1':u_1';\ \CEnd.
  \end{array}
\right) $ is not. Note, it is possible to transform choreographies
that are not strongly connected into strongly connected ones, by
adding extra interactions.

\smallskip

We are now ready to state our main theorem, which relates a
(projectable, strongly connected) choreography with its PRISM
projection.  In the sequel, $S_+$ denotes the extension of a global
state $S$ with the additional reserved variables $s_{\nrr}$ (one for
each module $\nrr$), introduced by the projection.  
\begin{thm}[Projection]\label{thm:epp}
  Let $C$ be a choreography with rates such that $\sconnected(C)$, let
  $\anno C = \mathsf{annot}(C,\sigma)$ for some fresh $\sigma$, and
  let $S_+$ be an extension of a global state $S$ such that for all
  $\nrr\in C$, we have that
  $S_+(s_{\nrr})=\topannotation\nrr {\anno C}$.  Then,
  \[
    (S, \anno C) \red{\lambda} (S', \anno {C}_1) \quad\text{if and only if}\quad
    \mathcal N(\anno C)\ \vdash\ S_+ \red{\lambda}
    S_+' ,
  \]
  such that: 
  \begin{itemize}
  \item for all $\nrr\in \anno C$, we have that
    $\projD(\nrr, \anno C_1) \subseteq \projD(\nrr, \anno C)$
  \item $S'=S'_{+}\backslash \{s_{\nrr}\}_{\nrr\in \anno C}$
  \item for all $\nrr\in \anno C$, we have that
    $S_+'(s_{\nrr}) = \topannotation\nrr {\anno C_1}$
  \end{itemize}
\end{thm}
\begin{proof}
  The proof proceeds by induction on $C$. We remove annotations when
  not relevant, in order to make the presentation clearer.
  \begin{itemize}

  \item
    $\anno
    C=\interactl{\nr^\iota}a{\nr_1^{\iota_1},\ldots,\nr_n^{\iota_n}}$. By
    the only applicable rule \textsf{(Interact)}, we have
    \[
      (S,
      \interactl{\nr^\iota}a{\nr_1^{\iota_1},\ldots,\nr_n^{\iota_n}})
      \red{\lambda_k} (S[u_k], C_k).
    \]
    By definition of projection, we obtain the following PRISM
    commands.  Module \(\nr\) is projected as:
    \[
      \left\{\commandBase {a_j} {\ s_{\nr}\!=\! \iota} {\lambda_j:}
      \ s_{\nr}'\!=\! \mathsf{top}_{\nr}(\anno{C}_j)\
      \&\ \projE {u_j}\nr\right\}_{j\in J}
      \,\cup\, \bigcup_{j\in J} \proj (\nr, \anno{C}_j)
    \]
    Modules \(\nr_i\) are projected as:
    \[
      \left\{\commandBase {a_j} {\ s_{\nr_i}\!=\! \iota_i} {1:\ s_{\nr_i}'\!=\!
      \mathsf{top}_{\nr_i}(\anno{C}_j)
      }
      \ \&\ \projE {u_j}{\nr_i}\right\}_{j\in J}
      \,\cup\, \bigcup_{j\in J} \proj (\nr_i, \anno{C}_j)
    \]
    Any other role $\nrr$ is projected as:
    \[
      \bigcup_{j\in J} \proj(\nrr, \anno C_j)
    \]
    We must show two properties: first, that the projection above can
    make the same transition; and second, that if the projection makes
    a transition, it necessarily corresponds to the transition of the
    choreography above.
    Observe that, since our annotated choreography is obtained from
    $\mathsf{annot}$, the state of the generated CTMC is uniquely
    identified by the labels $\iota$, $\iota_1$, \dots, $\iota_n$
    which, by strong connectedness, can only synchronise on one of the
    labels $a_j$.

    Therefore, the only-if direction of the theorem follows directly
    by applying rule \textsf{(Transition)}, whose premises are satisfied
    by deriving, via rule \textsf{(P$_2$)}, the command:
    \[
      \begin{array}{l}
        \commandBase{a_k}{(s_{\nr}=\iota) \land (s_{\nr_1}=\iota_1)
        \land \dots \land (s_{\nr_n}=\iota_n)}\\[1mm]
        \qquad
        \begin{array}{ll}
          (\lambda_k \cdot 1^n):\ &
                                    (\projE{u_k}{\nr})\ \&\
                                    (\projE{u_k}{\nr_1})\ \&\
                                    \dots\ \&\
                                    (\projE{u_k}{\nr_n})
          \\[1mm]
                                  &\ \&\
                                    (s_{\nr}'=\topannotation{\nr}{\anno C_k})\ \&\
                                    (s_{\nr_1}'=\topannotation{\nr_1}{\anno C_k})\ \&\
                                    \dots\ \&\
                                    (s_{\nr_n}'=\topannotation{\nr_n}{\anno C_k})
        \end{array}
      \end{array}
    \]
    \noindent
    For the converse direction, we must show that the above is the
    only enabled reduction in the PRISM semantics. As for modules
    $\nr$, $\nr_1$, \ldots, $\nr_n$, since the state $S_+$ yields
    $\iota$, $\iota_1$, \dots, $\iota_n$, no other reduction is
    possible. Other modules, projected as
    $\bigcup_{j\in J} \proj(\nrr, \anno C_j)$, are locally enabled by
    state $S_+$ to perform their next action. If, by contradiction, we
    can apply rule $\mathsf{(Transition)}$, derived by an application
    of rule $\mathsf{(P_1)}$ on some label $b$ (which must be
    different from $a$), on a different set of modules $\{\nrrr_k\}$,
    then we could show that there is no causal dependency between
    modules $\nr$, $\nr_1$, \ldots, $\nr_n$ and modules $\{\nrrr_k\}$,
    violating indeed the fact that the choreography is strongly
    connected. This reasoning is standard for projection theorems like
    the one we are proving~\cite{CHY12}.

  \item $\anno C=\ifTEl {E}a{\nr}{\anno C_1}{\anno C_2}$.
      In this case, the projection of $\nr$ is
      \begin{equation*}
      \quad\left\{ 
      \begin{array}{lll}
        \commandBase {} {s_{\nrr}\!=\! \iota\ \&\ E}{\ 1: s'_{\nrr}\!=\!
        \mathsf{top}_{\nrr}(\anno C_1)},\\ 
        \commandBase {} {s_{\nrr}\!=\! \iota\ \&\ \mathsf{not}(E)}
        {\ 1: s'_{\nrr}\!=\! \mathsf{top}_{\nrr}(\anno C_2)}
      \end{array}
      \right\}
      \ \cup\ \proj (\nrr, \anno C_1)
      \ \cup\
      \proj (\nrr, \anno C_2)
      \end{equation*}
      while, for all other modules, we have
      $ \proj (\nrr, \anno C_1) \ \cup\ \proj (\nrr, \anno C_2) $.  In
      this case, module $\nr$ is enabled by the uniqueness of $\iota$,
      while all other modules are not simply because we assume that
      the choreography is strongly connected. Hence, any other
      synchronisation or if-then-else statement is blocked, as it must
      involve $\nr$.

    \item $\anno C=X^{a,\sigma}$. By definition, assuming
      $X\defrec C_X\in \defin$, it must be the case that, for all
      $\nrr\in C$:
      \[\projD (\nrr, X^{a,\sigma}) = \{
        \commandBase {a} {s_{\nrr}\!=\! \sigma(\nrr)}{\ 1:
          s'_{\nrr}\!=\! } \mathsf{top}_{\nrr}(\anno
        C_X)\}\qquad\cup\quad\bigcup_{(Y\defrec
          C_Y)\in\defin}\big(\proj (\nrr, C_Y)\big)\]
      In the only-if direction, we know that
      $(S, X^{a,\sigma}) \red{1} (S, \anno C_X)$. From the uniqueness
      of $a$, we can derive
      $\mathcal N(\anno C)\ \vdash\ S_+ \red{1}
      S_+'$ by applying rule (\textsf{M}) on each module, then rule
      (\textsf{P$_2$}), and finally rule (\textsf{Transition}).
      Moreover, we observe that the command
      $s'_{\nrr}\!=\!  \mathsf{top}_{\nrr}(\anno C_X)$ has no effect
      on the state except modifying the extended variables with the
      top annotation in $\anno C_X$ which, by assumption again, it is
      unique. Additionally, we observe that for all $\nrr\in \anno C$,
      we have that
      $\projD(\nrr, \anno C_1) \subseteq \projD(\nrr, \anno
      C)$. Finally, for all $\nrr\in C$, we have that
      $S_+'(s_{\nrr}) = \topannotation\nrr {\anno C_1}$.

      In the other direction, we just observe that, for all $\nrr$,
      the extended state variables $s_{\nrr}$ are all set to
      $\sigma(\nrr)$. Therefore, given uniqueness of annotations, the
      projected modules can only simulate the procedure call.

    \item $C=\CEnd^\sigma$. Immediate.\qedhere
    
  \end{itemize}
\end{proof}
\noindent
We observe that the theorem departs from standard projection results
for choreography languages.  In most settings, both choreographies and
endpoints are expressed in a process algebra in which control flow is
encoded syntactically, and correctness is shown by relating reductions
of programs.  PRISM, instead, is declarative: a module consists of a
fixed set of commands, and execution is driven solely by state changes
through the enabling of commands.
Our choreography language combines both aspects.  Although it
maintains an explicit runtime state for variables, its syntax
determines the next action to be executed.  The projection reconciles
this mismatch by encoding control-flow information into the state: the
variables $s_{\nrr}$ act as a distributed program counter, yielding
the extended state $S_+$.
Consequently, choreography reduction corresponds in the projection to
transitions on $S_+$ rather than to syntactic changes of PRISM
modules.

\begin{rem}[Projection theorem for DTMC]\label{rem:dtmc}
  Theorem~\ref{thm:epp} is stated for choreographies with rates,
  projected according to Definition~\ref{def:projCTMC}. For
  choreographies with probabilities, projected according to
  Definition~\ref{def:projDTMC}, the correspondence is identical for
  conditionals, recursive calls and termination, whose projections
  perform exactly the same transitions with probability~1. An
  interaction is instead simulated by exactly two consecutive
  transitions of the projection:
  \[
    S_+ \red{\lambda_j} S_+[s_{\nr}'=\iota_j] \red{1} S_+'
  \]
  where the first transition is the internal probabilistic choice of
  the initiator $\nr$, which selects branch $j$ and only updates the
  reserved variable $s_{\nr}$, leaving $S$ unchanged.  The second is
  the synchronisation on label $a_j$, which applies the updates $u_j$
  and advances the reserved variables of all participants as in
  Theorem~\ref{thm:epp}. By uniqueness of state labels, the
  intermediate state enables exactly one command, so the composite
  probability of simulating branch $j$ is
  $\lambda_j\cdot 1=\lambda_j$. Theorem~\ref{thm:epp} thus holds for
  choreographies with probabilities once each interaction step is
  matched with this two-step transition of its projection.

  We also note that no normalisation (see Section~\ref{sec:prism}) is ever
  required for networks projected according to
  Definition~\ref{def:projDTMC}: by uniqueness of state labels and
  strong connectedness, in every reachable non-terminal state exactly
  one (possibly synchronised) command is enabled, and its
  probabilities sum to~$1$,
  while every other generated command carries probability~$1$.
\end{rem}

\begin{rem}
  Our theorem induces an asymmetry: projecting after reducing a
  choreography may result in fewer commands than projecting first and
  then reducing. This is because PRISM execution never alters the set
  of commands.  Let us focus on an example where we compare the
  semantics of the choreography \(X^{d,\sigma'}\), obtained from the
  recursive definition in Example~\ref{example2}, with that of its
  projection from Example~\ref{example-proj}. Starting from a state
  \(S_0\) in which \(x=0\) and \(y=0\), and from its extension
  \(S_0^+\), where additionally \(s_{\nr}=\sigma'(\nr)\) and
  \(s_{\nrr}=\sigma'(\nrr)\), both the choreography and its projection
  from Equation~\ref{eq:1} have a single possible transition:
  \begin{displaymath}
    \begin{array}{lll}
      (S_0,\ X^{d,\sigma'})\red 1
      \left(S_0,\ \interactBasel{\nr^{\iota_1}}{a}{\nrr^{\iota_2}}
      \left\{
      \begin{array}{lll}
        \lambda_1 : (x'=1)\ \&\ (y'=2);\ X^{b,\sigma_1} \\[0.5mm]
        \lambda_2 : (x'=3)\ \&\ (y'=1);\ X^{c,\sigma_2}
      \end{array}
      \right.\right)
      \\\\
      \mathcal N(X^{d,\sigma'})\vdash S_0^{+}
      \red{1} S_0^{+}[s_{\nr}'=\iota_1][s_{\nrr}'=\iota_2]
    \end{array}
  \end{displaymath}
  Clearly, the projection of the unfolding is strictly contained in
  the projection of \(X^{d,\sigma'}\). This is because
  the commands
  \(\commandBase{d}{s_{\nr}=\sigma'(\nr)}{1:(s_{\nr}'=\iota_1)}\) and
  \(\commandBase{d}{s_{\nrr}=\sigma'(\nrr)}{1:(s_{\nrr}'=\iota_2)}\)
  are no longer generated.
  However, the PRISM model obtained by projecting the reductum
  exhibits the same behaviour as the original one.  Relating these two
  states would require a suitable notion of equivalence induced by the
  semantics of PRISM, which we leave as future work. We conjecture
  that, under an appropriate equivalence notion, such as probabilistic
  bisimulation, our theorem could be extended to capture this
  additional feature.
\end{rem}

\section{Benchmarking}\label{sec:bench}
\newmark{R2.4}\newid{R2.3}{To evaluate our approach, we implemented the projection
from choreographies to PRISM in Java, using ANTLR to parse our
language and automatically generate the corresponding PRISM
modules. This implementation follows the projection formally defined
in Section~\ref{sec:proj} (folding recursive calls into the preceding
command) and ensures that each benchmark presented
below is automatically produced from a single global choreography
specification. The tool thus serves as an executable validation of
our formal framework: it is designed so that the semantics of the
generated PRISM models coincide with those of the original
choreographies and provides the basis for the empirical analysis
reported in this section.}

In the following, we present an experimental evaluation of our
language.  The examples highlight two main points: \emph{(i)} the
choreographic representation is significantly more concise than the
corresponding PRISM code, and \emph{(ii)} \new{in all but one case,
PRISM exhibits equivalent behaviour on both the projected and
original models}.
\newmark{R2.18/R3.3}%
\new{In this context, ``equivalent behaviour'' refers to agreement on
the quantitative properties that the original case studies were
designed to analyse, such as reachability and time-bounded
probabilities or expected values.  Our projection targets a structured
fragment of PRISM and is not intended to reproduce the exact syntactic
form or the full range of nondeterminism present in some hand-written
models.  As a consequence, we do not aim for bisimilarity or semantic
equivalence at the model level.  Instead, we verify that the projected
and reference models lead to the same results for the relevant
benchmark properties, which is the criterion of equivalence we adopt
in this evaluation.}

We focus on six benchmarks: a modified version of the example reported in Section \ref{sec:example}, 
a simple peer-to-peer protocol, 
the Bitcoin Proof-of-Work protocol \cite{DBLP:journals/concurrency/BistarelliNGLMV23}, 
the Hybrid Casper protocol \cite{DBLP:journals/distribledger/GallettaLMV23}, 
a synchronous leader-election protocol \cite{IR90}, 
and a modelisation of the dining cryptographers \cite{Cha88}.
The generated PRISM files are available in our online repository
\cite{repository}.
\newmark{R2.19}%
\new{For four of the benchmarks, PRISM performs exact
probabilistic model checking; for the two larger protocols (Bitcoin
Proof-of-Work and Hybrid Casper), exact verification is infeasible
due to state-space size and we resort to PRISM's statistical
model-checking (simulation) engine.}

\paragraph{A Modified thinkteam Protocol.}
In this modified version of the thinkteam protocol introduced in the earlier sections, we extend the protocol to involve generalised interactions with possible many receivers. Specifically, the \codechor{CheckOut} process now communicates with two users simultaneously, \codechor{User1} and \codechor{User2} each tasked with performing distinct actions upon access to the file. 

 In the first branch, \codechor{User1} increments the variable \codechor{x} by 1, while \codechor{User2} decrements the variable \codechor{y} by 1. Conversely, in the second branch, the roles are reversed, with \codechor{User1} decrementing \codechor{x} and \codechor{User2} incrementing \codechor{y}.

 \begin{lstlisting}[style=chor-color,breaklines=true, postbreak=\mbox{\textcolor{red}{$\hookrightarrow$}\space},caption={Choreography for the Modified thinkteam Protocol},captionpos=b,label={ex1-chor}]
C0 := CheckOut $\rightarrow$ User1, User2 : (+["lambda*1"] " " "(x=x+1)"  "(y=y-1)"; C1
                                      +["lambda*1"]  " " "(x=x-1)"  "(y=y+1)";  C2)
C1 :=  CheckOut $\rightarrow$ User1, User2 : (+["mu*1"] ; C0)  
C2 :=  CheckOut $\rightarrow$ User1, User2 : (+["theta*1"] ; C1   +["theta*1"] ;  C2)
 \end{lstlisting}

 \noindent
 Part of the generated PRISM model is reported in Listing \ref{ex1-gen}. 
It presents the behaviour in a decentralised form: 
interactions are encoded through synchronising commands distributed across the relevant modules. 
As a result, understanding the overall flow requires considering 
how these module-local transitions align, 
rather than reading a single global specification as in the choreographic representation.

\pagebreak

\begin{lstlisting}[style=prism-color,caption={Part of the generated PRISM model for the Modified thinkteam Protocol},captionpos=b,label={ex1-gen}]
module CheckOut
   CheckOut_STATE : [0..2] init 0;
   [MMHOL]  (CheckOut_STATE=0) -> lambda :  (CheckOut_STATE'=1);
   [FFSFW]  (CheckOut_STATE=0) -> lambda :  (CheckOut_STATE'=2);
   [ULCFN]  (CheckOut_STATE=1) -> mu : (CheckOut_STATE'=0);
   [YHHWG]  (CheckOut_STATE=2) -> theta : (CheckOut_STATE'=1);
   [XWSAO]  (CheckOut_STATE=2) -> theta : (CheckOut_STATE'=2);
endmodule
$\ldots$
module User2
   User2_STATE : [0..2] init 0;
   [MMHOL]  (User2_STATE=0) -> 1 : (y'=y-1)&(User2_STATE'=1);
   [FFSFW]  (User2_STATE=0) -> 1 : (y'=y+1)&(User2_STATE'=2);
   [ULCFN]  (User2_STATE=1) -> 1 : (User2_STATE'=0);
   [YHHWG]  (User2_STATE=2) -> 1 : (User2_STATE'=1);
   [XWSAO]  (User2_STATE=2) -> 1 : (User2_STATE'=2);
endmodule
\end{lstlisting}

\paragraph{Simple Peer-To-Peer Protocol}
This case study describes a simple peer-to-peer protocol based on BitTorrent\footnote{\url{https://www.prismmodelchecker.org/casestudies/peer2peer.php}}. The model comprises a set of clients trying to download a file that has been partitioned into $K$ blocks. Initially, there is one client that has already obtained all of the blocks and $N$ additional clients with no blocks. Each client can download a block from any of the others but they can only attempt four concurrent downloads for each block. The code we analyze with $K=5$ and $N=4$ is reported in Listing \ref{ex2-code}.
\begin{lstlisting}[style=chor-color,caption={Choreography for the Peer-To-Peer Protocol},captionpos=b,label={ex2-code}]
PeerToPeer := Client[i] $\rightarrow$ Client[i]: (+["rate1*1"]  "(b[i]1'=1)"$\&\&$" " . PeerToPeer
                         +["rate2*1"]  "(b[i]2'=1)"$\&\&$" " . PeerToPeer
                         +["rate3*1"]  "(b[i]3'=1)"$\&\&$" " . PeerToPeer
                         +["rate4*1"]  "(b[i]4'=1)"$\&\&$" " . PeerToPeer
                         +["rate5*1"]  "(b[i]5'=1)"$\&\&$" " . PeerToPeer)
\end{lstlisting} 

\begin{lstlisting}[style=prism-color,caption={Part of the generated PRISM program for the Peer-To-Peer Protocol},captionpos=b,label={ex2-gen}]
module Client1
   Client1 : [0..1] init 0;
   b11 : [0..1]; 
   b12 : [0..1]; 
   b13 : [0..1]; 
   b14 : [0..1]; 
   b15 : [0..1]; 

   [] (Client1=0)  $\rightarrow$ rate1 : (b11'=1)$\&$(Client1'=0); 
   [] (Client1=0)  $\rightarrow$ rate2 : (b12'=1)$\&$(Client1'=0); 
   [] (Client1=0)  $\rightarrow$ rate3 : (b13'=1)$\&$(Client1'=0); 
   [] (Client1=0)  $\rightarrow$ rate4 : (b14'=1)$\&$(Client1'=0); 
   [] (Client1=0)  $\rightarrow$ rate5 : (b15'=1)$\&$(Client1'=0); 
endmodule
   \end{lstlisting}
Part of the generated PRISM code is shown in Listing \ref{ex2-gen} and it is faithful with what is reported in the PRISM documentation. 
In Figure \ref{ex2-res}, we compare the probabilities that all clients (in a model with 4 
clients: \codeprism{Client1}, $\ldots$, \codeprism{Client4}) have received all blocks 
within the time interval $0 \leq T \leq 1.5$, as obtained from both our generated model 
and the model reported in the documentation. This property serves as a benchmark to 
evaluate whether the generated model preserves the expected behaviour of the original 
specification. In this case, there are no differences in the results or the time required 
to compute the property.

\begin{figure}[ht]
   \centering
   \includegraphics[scale=0.3]{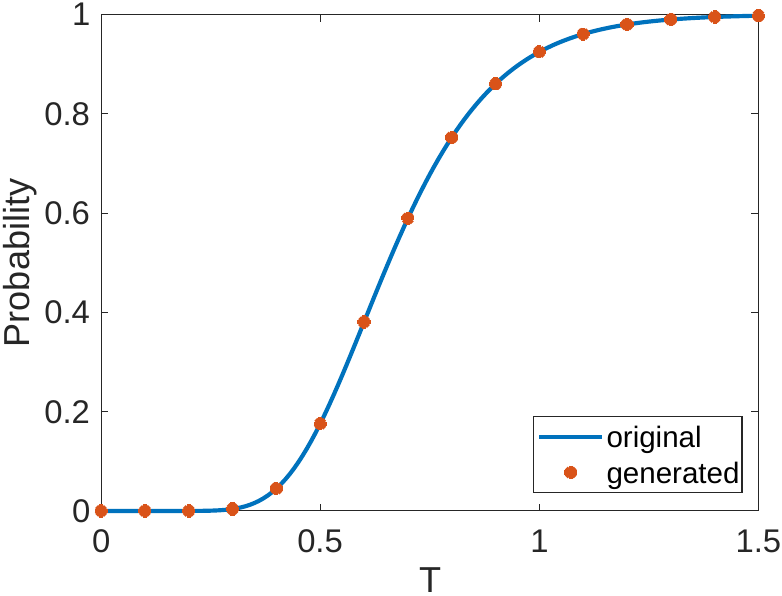}	
   \caption{Probability that clients received all the blocks before time $T$, with $0\leq T \leq 1.5$}
   \label{ex2-res}
   \end{figure}

\paragraph{Proof of Work Bitcoin Protocol.}
In \cite{DBLP:journals/concurrency/BistarelliNGLMV23}, the authors extended the PRISM model checker syntax to incorporate dynamic data types, enhancing its capabilities to model the Proof of Work protocol used in the Bitcoin blockchain \cite{bitcoin}. 

  \begin{lstlisting}[style=chor-color,breaklines=true, postbreak=\mbox{\textcolor{red}{$\hookrightarrow$}\space},caption={Choreography for the Proof of Work Bitcoin Protocol},captionpos=b,label={ex3-code}]
PoW $\coloneqq$ Hasher[i] $\rightarrow$ Miner[i] :
(+["mR*hR[i]"]  " " "(b[i]'=createB(b[i],B[i],c[i]))&(c[i]'=c[i]+1)" ; 
   Miner[i] $\rightarrow$ Network : (["rB*1"] "(B[i]'=addBlock(B[i],b[i]))" 
                                    foreach(k!=i) "(set[k]'=addBlockSet(set[k],b[i]))"@Network;PoW)
 +["lR*hR[i]"] ; 
   if "!isEmpty(set[i])"@Miner[i] then { 
      ["r"] "(b[i]'=extractBlock(set[i]))"@Miner[i] ;  
         Miner[i] $\rightarrow$ Network : (["1*1"] "(setMiner[i]'=addBlockSet(setMiner[i],b[i]))" 
                                             "(set[i]' = removeBlock(set[i],b[i]))";PoW) 
   }
   else{
      if "canBeInserted(B[i],b[i])"@Miner[i] then { 
         ["1"] "(B[i]'=addBlock(B[i],b[i]))&(setMiner[i]'=removeBlock(setMiner[i],b[i]))"@Miner[i];
         PoW 
      }
      else{PoW}
   })
   \end{lstlisting}

   \noindent
   In summary, the code depicts miners engaging in
   solving PoW, updating their ledgers, and communicating with the
   network.  The indices $i$ represent the module renaming feature of
   the choreographic language. Thus, each interaction will be repeated
   for each miner and hasher of the protocol that we are
   considering. The protocol works as follows. 
   When synchronising with
   the hasher, a miner tries to solve a cryptographic
   puzzle. Successful attempts add a new block to its ledger and update other miners' block sets. Unsuccessful attempts involve extracting a block, updating its ledger and block sets, and continuing the PoW process.

  \begin{figure}[ht]
   \centering
   \includegraphics[scale=0.35]{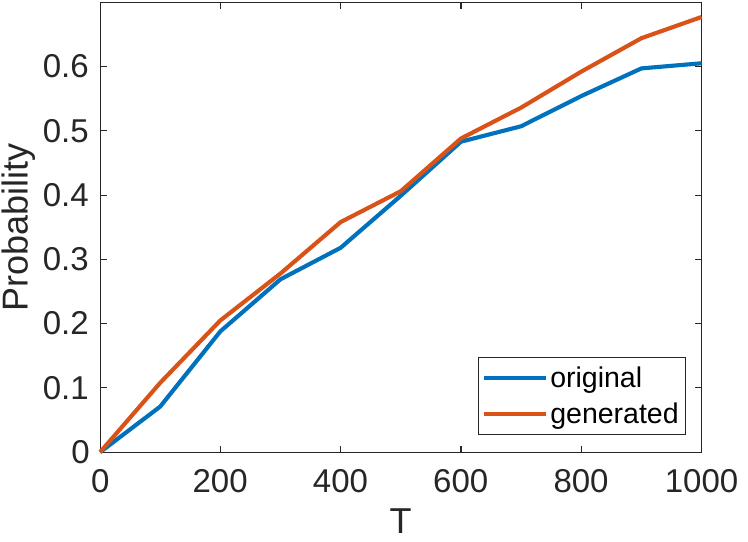}	
   \caption{Probability that a block is created within $T$ time units, $0\leq T\leq 1000$}
   \label{ex3-res}
\end{figure}
\noindent
The PRISM model we created is more verbose than the one in \cite{DBLP:journals/concurrency/BistarelliNGLMV23}, mainly because we consistently generate the else branch for if-then-else expressions, resulting in a higher number of instructions. Despite this, the experimental results for block creation probability within a bound time $T$ (Figure \ref{ex3-res}) remain close. Any discrepancies between the original and generated models are due to inherent variations in the simulation-based calculation of probability.

\paragraph{Hybrid Casper Protocol.}
We now present the Hybrid Casper Protocol \cite{DBLP:journals/distribledger/GallettaLMV23}. The Hybrid Casper protocol represents a hybrid consensus protocol for blockchains, merging features from both Proof of Work and Proof of Stake protocols. 
\begin{lstlisting}[style=chor-color,tabsize=2,breaklines=true, postbreak=\mbox{\textcolor{red}{$\hookrightarrow$}\space},	caption={Excerpt of choreography for the Hybrid Casper Protocol},captionpos=b,label={ex5-code}]
PoS := Hasher[i] -> Validator[i] :
$\quad\qquad$(+["mR*1"]  "(b[i]'=createB(b[i],L[i],c[i]))&(c[i]'=c[i]+1)"; 
$\quad\qquad\quad$if "!(mod(getHeight(b[i]),EpochSize)=0)"@Validator[i] then{$\ldots$}
$\quad\qquad\quad$else{
   $\quad\qquad\quad\quad$Validator[i] -> Vote_Manager :(["1*1"]  "(Votes'=addVote(Votes,b[i],stake[i]))"; PoS)}
$\quad\qquad$+["hR*1"]  ; if "!isEmpty(set[i])"@Validator[i] then { $\dots$ }
                     else{ PoS }
$\quad\qquad$+["rC*1"] "(lastCheck[i]'=extractCheckpoint(listCheckpoints[i],lastCheck[i]))"$\ldots$
\end{lstlisting}
The modelling approach is very similar to the one used for the Proof of Work Bitcoin protocol. Specifically, the Hybrid Casper protocol is represented in PRISM as the parallel composition of $n$ \codechor{Validator} modules, along with the modules \codechor{Vote\_Manager} and \codechor{Network}. Each \codechor{Validator} module closely resembles the \codechor{Miner} module from the previous protocol. The module \codechor{Vote\_Manager} is responsible for storing maps containing votes for each block and computing associated rewards/penalties.

The choreographic model for this example is reported in Listing \ref{ex5-code}. 
The code resembles that of the Proof of Work protocol, but each validator can either create a new block, receive blocks from the network module, or determine if it is eligible to vote for specific blocks.
For lack of space, we detailed only part of the code, the complete model can be found in \cite{repository}.

\begin{figure}[ht]
	\centering
	\includegraphics[scale=0.35]{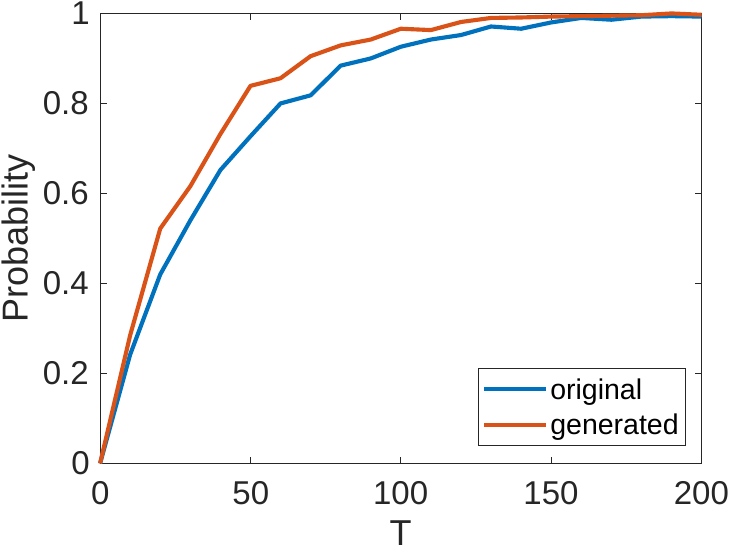}	
	\caption{Probability that a block is created within $T$ time units, $0\leq T\leq 200$}
	\label{ex5-res}
	\end{figure}
        \noindent
        The generated code is very similar the one outlined in
        \cite{DBLP:journals/distribledger/GallettaLMV23}, with the
        main distinction being the greater number of lines in our
        generated model.  This difference is due to the fact that
        certain commands could be combined, but our generation lacks
        the automatic capability to perform this check. While the
        results obtained for the probability of creating a block within the time $T$
        reported in Figure \ref{ex5-res} exhibit similarity, running
        simulations for the generated model takes PRISM 39.016
        seconds, compared to the 22.051 seconds required for the
        original model.

\paragraph{Synchronous Leader Election.}
This case study examines the synchronous leader election protocol proposed by Itai $\&$ Rodeh~\cite{IR90}, designed to elect a leader in a ring of $N$ processors by exchanging messages. The protocol operates in rounds, where each processor selects a random ID from $\{1, \ldots, K\}$, circulates it around the ring, and determines if a unique maximum ID exists. If so, the processor with this ID becomes the leader; otherwise, the process repeats in the next round.

For illustration, we considered the case where $ N=4 $ and $ K=8 $, following the PRISM model\footnote{\url{https://www.prismmodelchecker.org/casestudies/synchronous_leader.php}}. We modeled this example in our choreographic language, as shown in Listing~\ref{leader-code}, capturing the protocol's behaviour and dynamics.

\begin{lstlisting}[style=chor-color,caption={Choreography for the Synchronous Leader Election Protocol},captionpos=b,label={leader-code}]
Election := allSynch{ j in [1...4]
                  Process[j] : (true -> "1/K" : "(p[i]'=0)&(v[i]'=0)&(u[i]'=true)"+ 
                              $\ldots$ + "1/K" : "(p[i]'=7)&(v[i]'=7)&(u[i]'=true)") }.
   allSynch{
      Counter : ("(c<N-1)" -> "1" : "(c'=c+1)")
      Counter : ("(c=N-1)" -> "1" : "(c'=c)")
      Process1 : ("u1&!(p1=v2)&(c<N-1)" -> "1" : "(u1'=true)&(v1'=v2)")
      Process1 : ("u1&(p1=v2)&(c<N-1)" -> "1" : "(u1'=false)&(v1'=v2)&(p1'=0)")
      Process1 : ("!u1&(c<N-1)" -> "1" : "(u1'=false)&(v1'=v2)")
      Process1 : ("u1&!(p1=v2)&(c=N-1)" -> "1" : "(u1'=true)&(v1'=0)&(p1'=0)")
      Process1 : ("u1&(p1=v2)&(c=N-1)" -> "1" : "(u1'=false)&(v1'=0)&(p1'=0)")
      Process1 : ("!u1&(c=N-1)" -> "1" : "(u1'=false)&(v1'=0)")
      $\ldots$
   }.
   if "u1 | u2 | u3 | u4"@Counter then {
      Counter -> Process[i] : (["1*1"] "(c'=c)" "(u[i]'=false)&(v[i]'=0)&(p[i]'=0)". 
      allSynch {
         Counter : (true -> "1" : "(c'=c)")
         Process1 : (true -> "1" : " ")
         $\ldots$               
      } . END)
   }
   else{ Counter -> Process[i] : (["1*1"] "(c'=1)" "(u[i]'=false)&(v[i]'=0)&(p[i]'=0)"  . Election)}      
   \end{lstlisting} 
   While the generated model (Listing \ref{leader-prism}) successfully replicates the functionality of 
   the PRISM repository model, a key difference lies in the modular 
   structure of the two representations. 
   Specifically, the generated model adopts a simplified modular design by 
   grouping transitions more compactly in certain modules, such as the \codeprism{Counter} module. 
   This simplification reduces redundancy and may improve readability without altering the correctness or outcomes of the protocol. Importantly, this structural refinement does not impact the behaviour of the system, as the generated model remains functionally equivalent to the original PRISM repository model, as displayed in Figure \ref{leader-res}, where the two curves coincide for $L\geq 3$.
   \begin{figure}[H]
      \centering
      \includegraphics[scale=0.3]{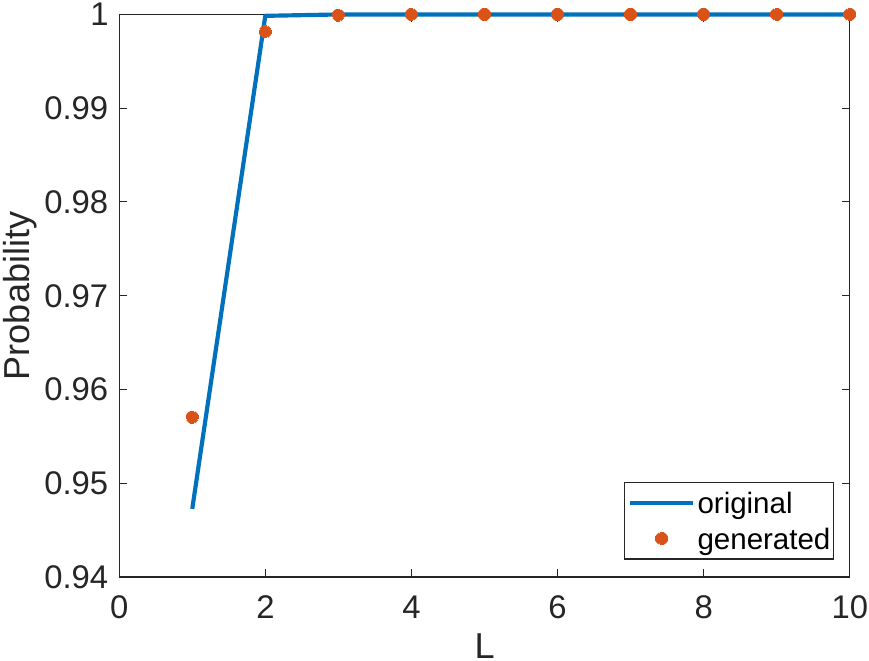}
      \caption{The probability of electing a leader within $L$ rounds, with $1 \leq L \leq 10$}
      \label{leader-res}
   \end{figure}
   \begin{lstlisting}[style=prism-color,caption={Part of the generated PRISM model for the Synchronous Leader Election Protocol},captionpos=b,label={leader-prism}]
module Counter
   Counter : [0..4] init 0;
   c : [0..N-1] init 0;
   [YQBDX] (Counter = 0)&(c<N-1) -> 1 : (c'=c+1)&(Counter'=1);
   [YQBDX] (Counter = 0)&(c=N-1) -> 1 : (c'=c)&(Counter'=1);
   [ELTMI] (Counter=1)&(u1 | u2 | u3 | u4) -> 1 : (c'=c)&(Counter'=2);
   [LJTIP] (Counter=1)&!(u1 | u2 | u3 | u4) -> 1 : (c'=1)&(Counter'=0);
   [AWUQP] (Counter = 2)-> 1 : (c'=c)&(Counter'=2);
endmodule
module Process1
Process1 : [0..4] init 0;
   p1 : [0..K-1] init 0;
   v1 : [0..K-1] init 0;
   u1 : bool;
   [BKKXT](Process1 = 0) ->  1/K:(p1'=0)&(v1'=0)&(u1'=true)&(Process1'=1) 
                              + 1/K:(p1'=1)&(v1'=1)&(u1'=true)&(Process1'=1) 
                              + 1/K:(p1'=2)&(v1'=2)&(u1'=true)&(Process1'=1) 
                              + 1/K:(p1'=3)&(v1'=3)&(u1'=true)&(Process1'=1) 
                              + 1/K:(p1'=4)&(v1'=4)&(u1'=true)&(Process1'=1) 
                              + 1/K:(p1'=5)&(v1'=5)&(u1'=true)&(Process1'=1) 
                              + 1/K:(p1'=6)&(v1'=6)&(u1'=true)&(Process1'=1) 
                              + 1/K:(p1'=7)&(v1'=7)&(u1'=true)&(Process1'=1);
   [YQBDX] (Process1 = 1)&u1&!(p1=v2)&(c<N-1) -> 1 : (u1'=true)&(v1'=v2)&(Process1'=2);
   [YQBDX] (Process1 = 1)&u1&(p1=v2)&(c<N-1) -> 1 : (u1'=false)&(v1'=v2)&(p1'=0)&(Process1'=2);
   [YQBDX] (Process1 = 1)&!u1&(c<N-1) -> 1 : (u1'=false)&(v1'=v2)&(Process1'=2);
   [YQBDX] (Process1 = 1)&u1&!(p1=v2)&(c=N-1) -> 1 : (u1'=true)&(v1'=0)&(p1'=0)&(Process1'=2);
   [YQBDX] (Process1 = 1)&u1&(p1=v2)&(c=N-1) -> 1 : (u1'=false)&(v1'=0)&(p1'=0)&(Process1'=2);
   [YQBDX] (Process1 = 1)&!u1&(c=N-1) -> 1 : (u1'=false)&(v1'=0)&(Process1'=2);
   [ELTMI] (Process1=2) -> 1 : (u1'=false)&(v1'=0)&(p1'=0)&(Process1'=3);
   [LJTIP] (Process1=2) -> 1 : (u1'=false)&(v1'=0)&(p1'=0)&(Process1'=0);
   [AWUQP] (Process1 = 3) -> 1 :  (Process1'=4);
endmodule
$\ldots$
   \end{lstlisting}

\paragraph{Dining Cryptographers}
The generated model for this example does not faithfully model the original one. We chose 
to include it in the paper to demonstrate the limitations of our approach, 
specifically in cases where the abstraction may not fully capture the behaviour of the original protocol. 
This allows us to analyze and understand where discrepancies may arise, 
highlighting areas for improvement in the model generation process.
This case study explores the dining cryptographers protocol introduced by Chaum~\cite{Cha88}, which allows a group of $N$ cryptographers to determine whether their master has anonymously paid for dinner without revealing the identity of the payer. The protocol functions by having each cryptographer flip a fair coin and share the outcome with their right-hand neighbour. Each cryptographer then publicly declares whether the two coins they observe—one they flipped and one received from the left—match or differ. If a cryptographer is the payer, they deliberately alter their response. The final count of ``differ'' statements follows a predictable pattern: an odd count indicates that one of the cryptographers paid, while an even count means the master paid.

To illustrate the protocol, we examined the case where $N=3$, using the PRISM model\footnote{\url{https://www.prismmodelchecker.org/casestudies/dining_cryptographers.php}} reported in the official repository. We expressed this scenario in our choreographic language, as shown in Listing~\ref{dc-code}.
\begin{lstlisting}[style=chor-color,caption={Choreography for the Dining Cryptographers Protocol},captionpos=b,label={dc-code}]
Crypto := if "(coin[i]=0)"@crypt[i] then {
               crypt[i] -> crypt[i] : (+["0.5*1"] "(coin[i]'=1)" . Crypto2
                                          +["0.5*1"] "(coin[i]'=2)" . Crypto2) 
            }
            else{ Crypto }

Crypto2 := if "((coin[i]>0)&(coin[i+1]>0))"@crypt[i] then{
               if "(coin[i]=coin[i+1])"@crypt[i] then {
                  if "(pay=p[i])"@crypt[i] then { Crypto3 }
                  else{ ["1"] "(agree[i]'=1)"@crypt[i]. Crypto3 }
               }
               else{
                     if "(pay=p[i])"@crypt[i] then { ["1"] "(agree[i]'=1)"@crypt[i]. Crypto3 }
                     else{ Crypto3 }
               }
            }
            else { Crypto2 }

Crypto3 := allSynch{ j in [1...3]
               crypt[j] : (true -> "1" : "true;" )
            }.END
\end{lstlisting}
This code defines a choreography model using three distinct choreographies (\codechor{Crypto}, \codechor{Crypto2}, and \codechor{Crypto3}) to avoid redundancy and optimise the code structure. By using separate choreographies, we can reuse common logic without rewriting the entire code.
The initial probabilistic branching demonstrates the use of recursion, where both branches of the probabilistic transition ultimately lead to the same state. 

In fact, in the generated PRISM model in Listing~\ref{dc-code-prism}, we can observe that from the initial state \codeprism{crypt1 = 0}, we transition to two possible branches where \codeprism{crypt1 = 1} regardless of whether coin1 takes the value 1 or 2. This is a result of the recursion in the choreography, where both branches follow the same recursive path that leads to \codeprism{crypt1 = 1}. The transitions in the model are shown in the following PRISM code, where both transitions have a 50$\%$ probability of either setting \codeprism{coin1 = 1} or \codeprism{coin1 = 2}, and both lead to the same updated state \codeprism{crypt1 = 1}. Note that the code provided here only shows a part of the generated PRISM model. The code for \codeprism{crypt2} and \codeprism{crypt3} follows the same structure.

\begin{lstlisting}[style=prism-color,caption={Part of the generated PRISM model for the Dining Cryptographers Protocol},captionpos=b,label={dc-code-prism},escapechar=|]
$\ldots$
module crypt1
crypt1 : [0..2] init 0;
coin1 : [0..2] init 0;
s1 : [0..1] init 0;
agree1 : [0..1] init 0;
[] (crypt1=0)&((coin1=0)) -> 0.5 : (coin1'=1)&(crypt1'=1)+0.5 : (coin1'=2)&(crypt1'=1); |\label{first-line-crypto}|
[] (crypt1=0)&!((coin1=0)) -> 1:(crypt1'=0);
[] (crypt1=1)&(coin1>0)&(coin2>0)&((coin1=coin2))&!((pay=p1)) -> 1:(agree1'=1)&(crypt1'=2);
[] (crypt1=1)&(coin1>0)&(coin2>0)&((coin1=coin2))&((pay=p1)) -> 1:(crypt1'=2);
[] (crypt1=1)&(coin1>0)&(coin2>0)&!((coin1=coin2))&((pay=p1)) -> 1:(agree1'=1)&(crypt1'=2);
[] (crypt1=1)&(coin1>0)&(coin2>0)&!((coin1=coin2))&!((pay=p1))-> 1:(crypt1'=2);
[LERZX] (crypt1 = 2) -> 1:(crypt1'=2);
endmodule
$\ldots$
\end{lstlisting}

\noindent
In this case, however, the generated model does not exhibit the same behaviour as the
original one in the PRISM repository. When analysing the probability of anonymity, the
result obtained by the original model, as reported on the website, is \(\,0.25\,\), whereas in
our generated model the probability is \(0\). This discrepancy arises from a structural
difference in how transitions are enabled in the two models. In the original PRISM model,
the initial coin toss of cryptographer~1 is specified as
\begin{lstlisting}[style=prism-color, frame=none, numbers=none, xleftmargin=0pt]
[] (coin1 = 0) -> 0.5 : (coin1' = 1) & (crypt1' = 1) + 0.5 : (coin1' = 2) & (crypt1' = 1);
\end{lstlisting}
and is enabled solely on the condition \texttt{coin1 = 0},
independently of the current value of \texttt{crypt1}.
\newmark{R1.15}%
\new{In contrast, in our generated model the corresponding transition
is guarded by both \texttt{coin1 = 0} and \texttt{crypt1 = 0} (see
line~\ref{first-line-crypto} of Listing~\ref{dc-code-prism}). This
additional guard is not incidental: it reflects a fundamental property
of our projection, namely that every interaction in a choreography is
associated with a unique control state of each role. The projection
introduces control variables to ensure that, for each module, only
the commands corresponding to the current control state are
enabled. As a result, our translated models enforce a single enabled
action per module per state, whereas the original PRISM model permits
the initial coin toss to be enabled in several control contexts.}

\new{This mismatch in the enabling conditions reduces the reachable
state space of the generated model and leads to different quantitative
results (95 states and 194 transitions in our model versus 380 states
and 776 transitions in the original one). More generally, this
illustrates a limitation of our current projection approach: PRISM
models in which certain probabilistic transitions are enabled
independently of the logical control flow of a process -- potentially
coexisting with other enabled commands -- cannot be reproduced exactly
via projection from a choreography, because the projection enforces a
control discipline absent in such models.}

\paragraph{Code Size Comparison}
\newmark{R2.3}%

\new{To complement the qualitative discussion of the benchmarks, we
provide a quantitative comparison of the size of each choreography
and the corresponding generated PRISM models. Table~\ref{tab:codesize}
summarises the number of lines of code (LoC) of each choreography and
of the corresponding generated PRISM model, excluding variable
declarations.}

\begin{table}[h]
\centering

\begin{tabular}{lccc}
\toprule
\textbf{Benchmark} & \textbf{Choreography LoC} & \textbf{PRISM LoC}  \\
\midrule
Modified thinkteam protocol & 4  & $\sim$15  \\
Peer-to-Peer protocol       & 6  & $\sim$20       \\
Bitcoin Proof-of-Work       & $\sim$45 & $\sim$100 \\
Hybrid Casper               & $\sim$70 & $>$220   \\
Leader Election             & $\sim$20 & $\sim$60 \\
Dining Cryptographers       & $\sim$17 & $>$40 \\
\bottomrule
\end{tabular}
\caption{Size comparison between choreographies and generated PRISM models.}
\label{tab:codesize}
\end{table}

\noindent
\new{The table highlights a structural difference between the two
formalisms: choreographies describe global behaviour in a single
specification, whereas PRISM distributes behaviour across several
modules using synchronising commands. As a result, choreography
descriptions tend to be more compact, while the generated PRISM
models are necessarily more verbose in order to express the
equivalent endpoint behaviour. This comparison motivates the use of
choreographies as a higher-level modelling notation, while still
enabling automated generation of analysable PRISM code.}

\section{Related Work and Discussion}\label{sec:relwork}
\mypar{Related Work.}  Choreographic programming~\cite{M23} is a
language paradigm for specifying the expected interactions
(communications) of a distributed system from a global viewpoint,
from which decentralised implementations can be generated via
projection. The notion of choreography has been substantially explored
in the last decade, both from a theoretical perspective,
e.g.,~\cite{CHY12,CM13}, to full integration into fully-fledged
programming languages, such as WS-CDL~\cite{HYC07b} and
Choral~\cite{GMP24}. Nevertheless, there is a scarcity of research on
probabilistic aspects of choreographic programming.
To the best of our knowledge, Aman and Ciobanu~\cite{AC19,AC22} are
the only ones who studied the concept of choreography and
probabilities. Their work augments multiparty session types (type
abstractions for communicating systems that use the concept of
choreography) with a probabilistic internal choice similar to the one
used by our choreographic branching. However, they do not provide any
semantics with state in terms of Markov chains, and, most importantly,
they do not project into a probabilistic declarative language model
such as PRISM.
Carbone et al.~\cite{CGHL10} define a logic for expressing properties
of a session-typed choreography language. However, the logic is
undecidable and has no model-checking algorithm. 
As far as our knowledge extends, there is currently no work that
generates probabilistic models from choreographic languages that can
be then model-checked.

\newid{R3.2}{A complementary line of research concerns the implementability and
  realisability of global specifications, where the central question
  is whether a given global transition system can be decomposed into
  synchronously communicating local components whose composition
  preserves the intended behaviour \cite{10.1007/3-540-46691-6_17,
    Mukund2002, 10.1007/978-3-031-47963-2_15,
    10.1007/978-3-031-62697-5_10}.  In those settings, the global
  description is typically unconstrained and substantial effort goes
  into characterising which behaviours admit a distributed
  realisation.  By contrast, choreographies impose a structured form
  of global specification in which all interactions are explicitly
  declared.  This discipline ensures that every choreography is
  directly realisable via projection, but it also restricts the kinds
  of distributed behaviours that can be expressed, placing our
  approach within a well-behaved subspace of the broader realisability
  landscape.}

\mypar{Discussion and Future Work.}
The ultimate goal of the proposed framework is to use the concept of
choreographic programming to improve several aspects, including
usability, correctness, and efficiency in modelling and analysing
systems.
In this paper, we address usability and efficiency of modelling
systems by proposing a probabilistic choreography language.
Our language improves the intuitive modelling of concurrent
probabilistic systems. Traditional modelling languages often lack the
expressive clarity needed to effectively capture the intricacies of
such systems. By designing a language specific for choreographing
system behaviours, we provide an intuitive means of specifying system
dynamics. This approach enables a more natural and straightforward
modelling process, essential for accurately representing real-world
systems and ensuring the efficacy of subsequent analysis.
Although choreographies and the projection function aim to abstract
away low-level details, providing instead a higher-level
representation of system behaviours, the choreographic approach
naturally imposes some restrictions on expressivity.  \newid{R3.1}{A
  choreography prescribes a single global interaction structure that
  determines, for each role, a unique enabled action at any control
  state.  This discipline is crucial for correctness-by-construction
  but also limits the class of systems that can be expressed: models
  that rely on multiple independently enabled transitions,
  nondeterministic interleavings, or asynchronous local evolution
  cannot, in general, be captured by our current framework.  The
  systems supported by our approach are therefore those whose
  behaviour can be organised around explicitly coordinated
  interactions with a globally determined branching structure.}  Some of the case studies presented in the
PRISM documentation~\cite{PRISMdoc} cannot be modeled by using our
current language. Specifically, there are two main cases where our
approach encounters limitations:
\textit{(i)} in the asynchronous leader election case study, our
language prohibits the use of an $\lq\lq$if-then$\lq\lq$ statement without an
accompanying $\lq\lq$else$\lq\lq$ to prevent deadlocked states;
\textit{(ii)} in probabilistic broadcast protocols or cyclic server
polling system models, the system requires probabilistic branching to
synchronise different modules based on the selected branch.
These issues could be fixed by extending our choreographic language
further and are therefore left as future work.

Additionally to these extensions, we conjecture that our semantics for
choreographies may be used for improving performance by directly
generating a CTMC or a DTMC from a choreography, bypassing the
projection into the PRISM language. In fact, the Markov chain
construction from a choreography seems to be faster than the
construction from a corresponding projection in the PRISM language,
since it is not necessary to take into account all the possible
synchronisations in the rules from Fig.~\ref{fig:semantics}. A formal
complexity analysis, an implementation, and performance benchmarking
are left as future work.

In conclusion, this paper has introduced a framework for addressing
the challenges of modelling and analysing concurrent probabilistic
systems.
The development of a choreographic language with tailored syntax and
semantics offers an intuitive modelling approach. We have established
the correctness of a projection function that translates choreographic
models to PRISM-compatible formats. Additionally, our compiler enables
seamless translation of choreographic models to PRISM, facilitating
powerful analysis while maintaining expressive clarity. These
contributions bridge the gap between high-level modelling and robust
analysis in probabilistic systems, paving the way for advancements in
the field.

\bibliographystyle{alphaurl}
\bibliography{biblio}

\end{document}